\documentclass[journal]{IEEEtran}
\usepackage{amsmath,amsfonts}
\usepackage{algorithmic}
\usepackage{algorithm}
\usepackage{array}
\usepackage[caption=false,font=normalsize,labelfont=sf,textfont=sf]{subfig}
\usepackage{textcomp}
\usepackage{stfloats}
\usepackage{url}
\usepackage{verbatim}
\usepackage{graphicx}
\usepackage{cite}
\usepackage{makecell}
\usepackage{booktabs}
\usepackage{multirow}
\usepackage{siunitx}
\usepackage{color} 
\usepackage{soul} 

\begin{document}

\title{SF-MMCN: Low-Power Sever Flow Multi-Mode Diffusion Model Accelerator}

\author{
Huan-Ke Hsu, 
I-Chyn Wei \\ {Chang Gung University, Taoyuan, Taiwan} \\ 
{~~} \\
T. Hui Teo \\ {Singapore University of Technology and Design, Singapore} \\
tthui@sutd.edu.sg
}




\maketitle

\begin{abstract}
Generative Artificial Intelligence (AI) has become incredibly popular in recent years, and the significance of traditional accelerators in dealing with large-scale parameters is urgent. With the diffusion model's parallel structure, the hardware design challenge has skyrocketed because of the multiple layers operating simultaneously. Convolution Neural Network (CNN) accelerators have been designed and developed rapidly, especially for high-speed inference. Often, CNN models with parallel structures are deployed. In these CNN accelerators, many Processing Elements (PE) are required to perform parallel computations, mainly the multiply and accumulation (MAC) operation, resulting in high power consumption and a large silicon area. In this work, a Server Flow Multi-Mode CNN Unit (SF-MMCN) is proposed to reduce the number of PE while improving the operation efficiency of the CNN accelerator. The pipelining technique is introduced into Server Flow to process parallel computations. The proposed SF-MMCN is implemented with TSMC 90-nm CMOS technology. It is evaluated with VGG-16, ResNet-18, and U-net. The evaluation results show that the proposed SF-MMCN can reduce the power consumption by 92\%, and the silicon area by 70\%, while improving the efficiency of operation by nearly 81 times. A new FoM, area efficiency (GOPs/$mm^2$) is also introduced to evaluate the performance of the accelerator in terms of the ratio throughput (GOPs) and silicon area ($mm^2$). In this FoM, SF-MMCN improves area efficiency by 18 times (18.42).

\end{abstract}


\section{Introduction}
\IEEEPARstart{W}{ith} the advanced of deep learning, the applications of Convolution Neural Network (CNN) have sprung up in recent years. The concept of CNN accelerators has emerged in a variety of architectures. However, due to the large number of parameters and the complexity of images skyrocketing with the development of deep learning, the requirement of low-power, high speed in CNN accelerator has been significant gradually. To reduce the scale of features in convolution computation, some studies~\cite{CarryapproFA, AppCons_LOA, AppCons_LOCA, AppAddDNN}, the proposed approximate convolution circuits, which directly decrease computation resources by capturing partial data from input feature and weight data, the complexity of convolution would decrease distinctly. Moreover, because of the small size of the input data, the critical path of multipliers and the full adders becomes shorter, which would save the area and power of the computation. And the timing cost by CNN can accelerate. Since the contents of a conventional multiplier are full adders, there have been many assembles of structures of multipliers nowadays, which have been the same as the approximate multiplier~\cite{PowerCarryPre, DRUM, SSM_DSM, BPE_CPE}. In other words, implementing the approximate multiplier by improving the basic logic unit of the conventional multiplier becomes the most intuitive method. There are three main features of traditional multipliers: 1) partial-product generation, 2) partial-product accumulation, and 3) final addition. Generally, one of the serious problems of multipliers is that it takes some time to accumulate the partial product and the final accumulate because it is limited by the hardware. Consequently, compressors ~\cite{Comp_basic, TwoComp4_2, LowPowerComp4_2} reduce the partial product and the critical path without any influence on circuit performances.

Although approximate computation units reduce the complexity of convolution, the accuracy lost has become another serious issue of CNN accelerators. Due to the application of image recognition, accuracy loss is one of the most fatal factors of CNN accelerators~\cite{Eyeriss, ReconvHA, Flexible, CARLA, OverFF, Pre-scalable, ECG, ISSCC, ISCAS, TCAD}. Therefore, researchers try to focus on the reconfigurable structure of CNN accelerators instead of modifying the algorithm of MAC computation. The CNN accelerator that was proposed in \cite{CARLA} can complete the convolution in different shapes of the kernel. The data flow of \cite{CARLA} precomputes the sub-output and stores it in memories to complete the reuse of the data in $3\times3$ convolution. When performing the $1\times1$ convolution, the architecture in \cite{CARLA} changed the dimension of the convolution from width to channel at a time. There are 195 PEs in \cite{CARLA} divided into 65 columns in the top structure, increasing the throughput of the circuit. However, since the PE array is allocated, the efficiency will decrease when the size of the input features or the weight data are not divided by 3 or 65.

Due to the bit-width in software being 64 bits, most studies adopt the fix-point precision technique to conduct hardware computation before the occurrence of quantization. However, the lost data from fix-point data and data truncation between layers result in accuracy loss. Quantization tries to reduce the bit-width dramatically, but the accuracy is likely to be less than the traditional method. Therefore, researchers adapt the quantization function into CNN accelerators. The implementation of this technique is exemplified in \cite{ISSCC}, which introduces the Quantized Network-Acceleration Processor (QNAP) structure. In this innovative design, weight data, upon undergoing quantization, can be seamlessly assembled with other quantified weight data. Consequently, the PE structure in \cite{ISSCC} departs from conventional designs. With the quantization of weight data, \cite{ISSCC} reduces power consumption. 

\begin{figure}[H]
\centering
\subfloat[]{
\includegraphics[width=0.3\textwidth]{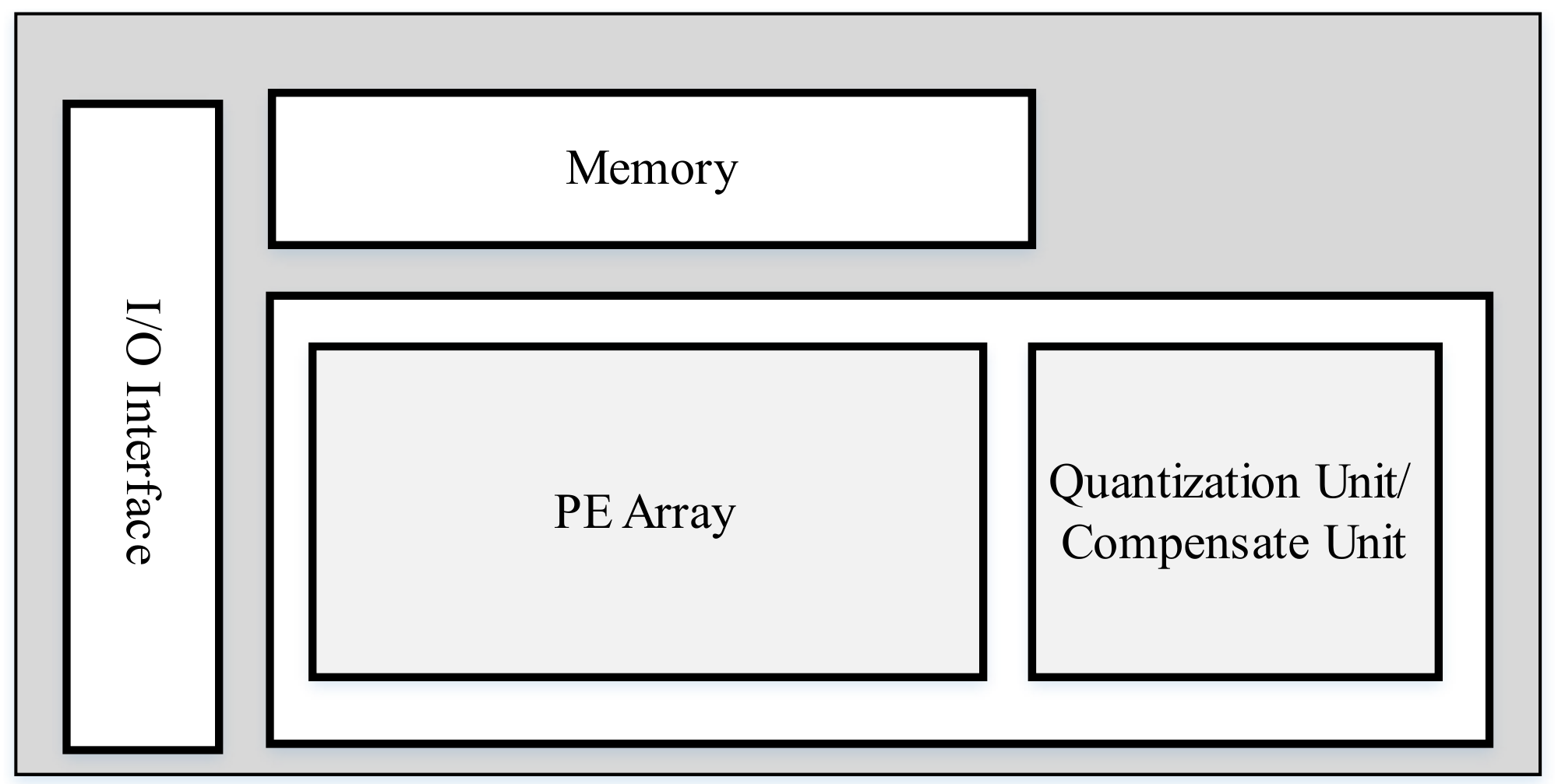}}
\hfill
\subfloat[]{
\includegraphics[width=0.25\textwidth]{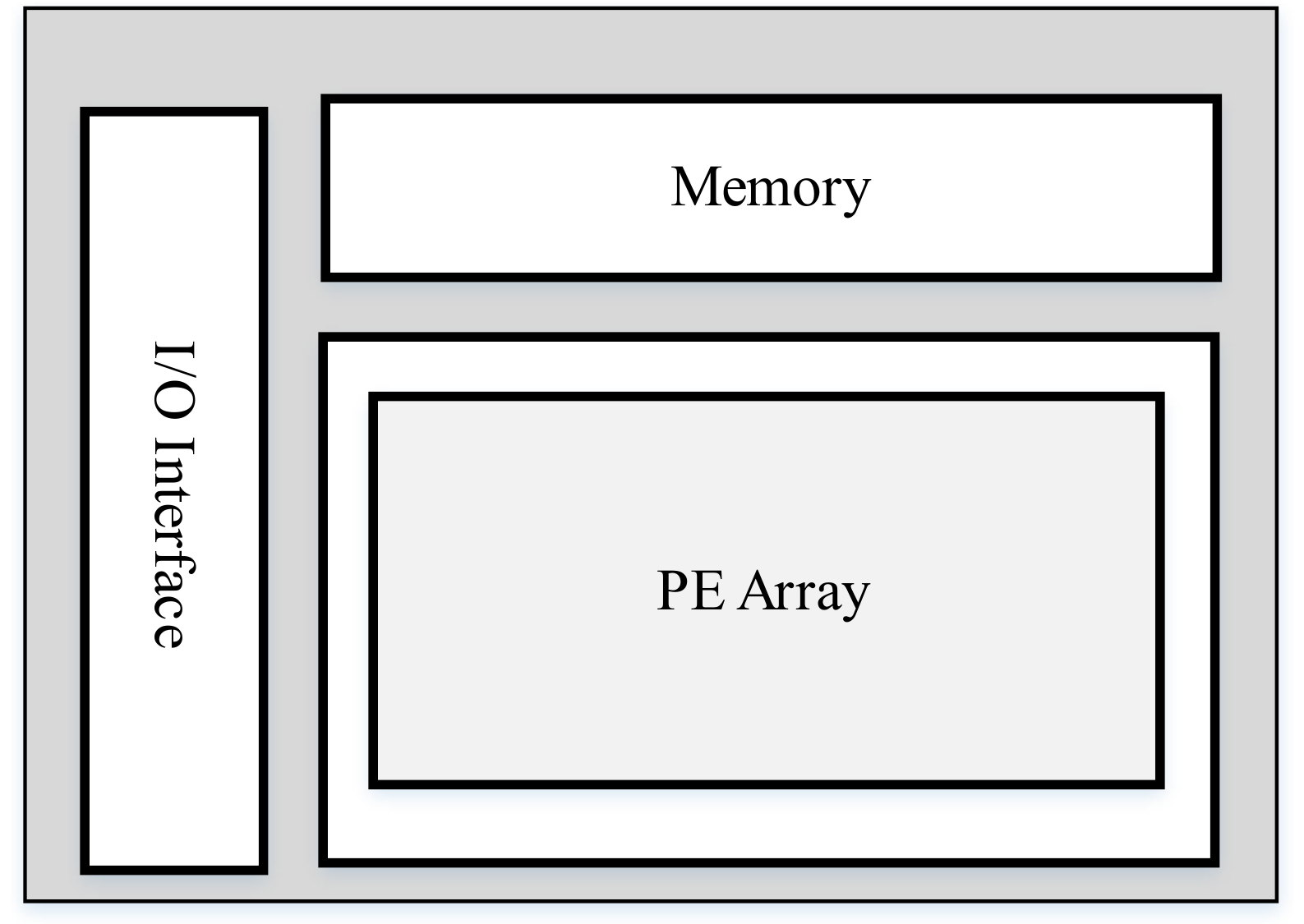}}
\caption{(a) A CNN accelerator structure with quantization technique. (b) A Traditional CNN accelerator structure.} \label{fig:QuantvsTradi}
\end{figure}

There are two CNN accelerator architectures between traditional structure and quantified structure, Figure~\ref{fig:QuantvsTradi}. CNN accelerators with quantization technique require a compensation or quantization unit of hardware implementation, as indicated in Figure~\ref{fig:QuantvsTradi} (a). Due to lower bit-width data after quantization, a smaller area is required of a PE array. On the other hand, the traditional structure cost a larger hardware area owing to the elimination of a compensate or quantization unit Figure~\ref{fig:QuantvsTradi} (b). 

Another challenge in CNN accelerators is that due to the rapid advancement of CNN models, there are lots of CNN models consisting of parallel structures such as residual and bottleneck blocks. \cite{CARLA}, \cite{ISSCC}, and \cite{ISCAS} implement the CNN accelerators with the ability to operate parallel CNN models. However, to address parallel CNN structure, there are two methods of CNN accelerators in recent studies, parallel, and series, both strategies result in different drawbacks of hardware implementations. The parallel method would utilize more hardware to complete MAC operation due to the large number of input data compared to the series method. However, the series strategy takes more time due to dividing residual blocks into several convolution layers. Therefore, it's significant to strike a balance or get rid of both shortcomings of the two strategies mentioned above. \par


Furthermore, given the trend of the diffusion model \cite{diffusion} for generative AI, U-net
\cite{unet}, as one of the most popular models for the de-noise diffusion model, consists of many parallel structures. Since generative AIs are responsible for dealing with enormous data. What's more, the applications of generative AI mostly require real-time characteristics such as translation and use as a personal secretary. Consequently, accelerating the AI model with a parallel structure has become the most necessary challenge in recent years. 

This paper proposed an enhanced version of \cite{my, hsu2024sfmmcn}, named Sever Flow Multi-Mode Convolution Neural Network Unit (SF-MMCN). The proposed SF-MMCN achieves high power efficiency and area efficiency under parallel CNN models. With an internal pipeline and SF structure, the proposed SF-MMCN can feast in a variety of environments and overcome drawbacks resulting from the series and parallel strategies discussed before. By adjusting the mode of hardware, the proposed structure supports not only the residual block but also the time parameter process in U-net. The proposed SF-MMCN contributes to a low power structure, small hardware area, and high operation efficiency.  

\section{Definitions and Background}
\begin{figure}[H]
\centering
\includegraphics[width=0.4\textwidth]{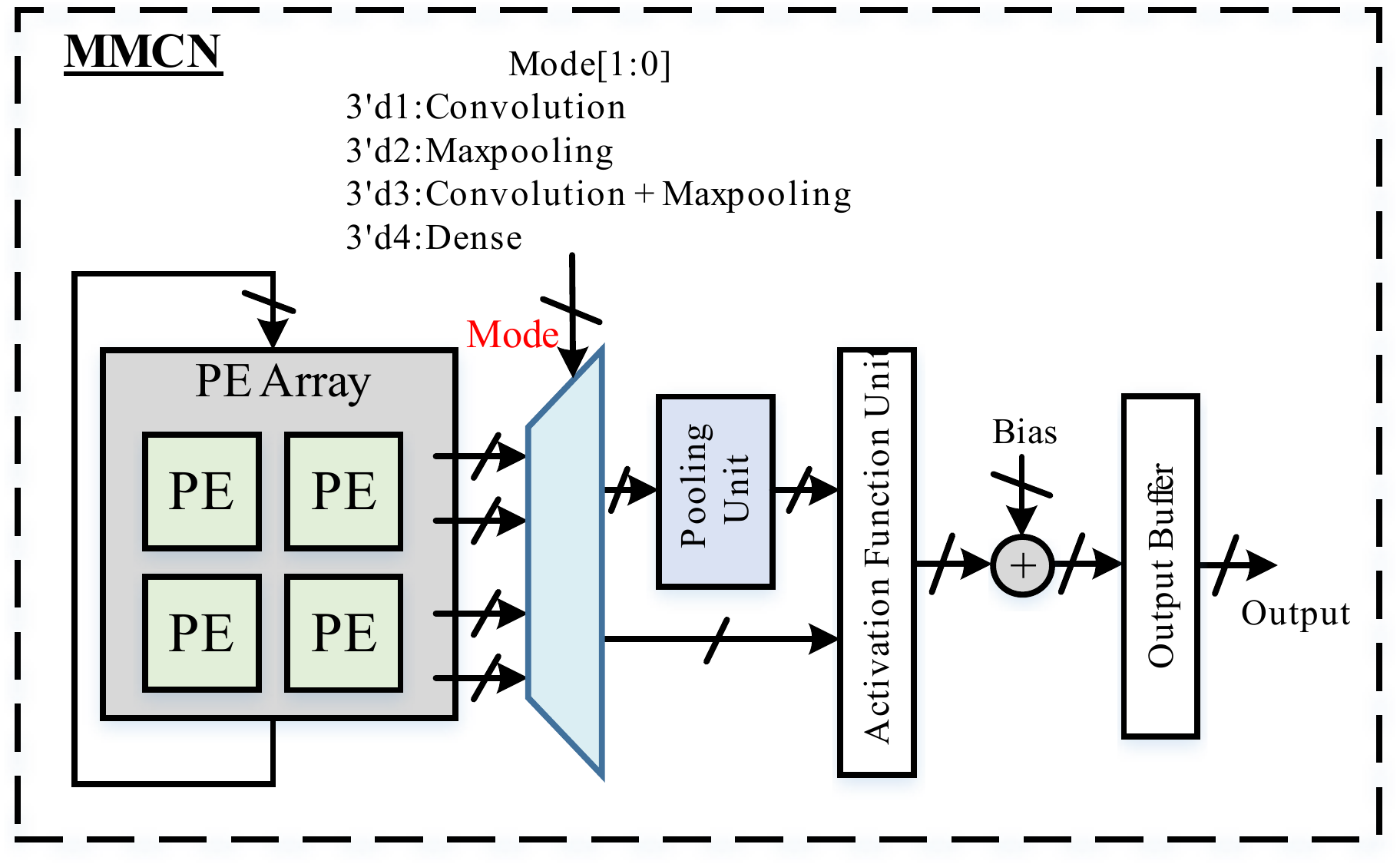}
\caption{MMCN structure in \cite{my}.} 
\label{fig:MMCN}
\end{figure}

As mentioned above, this paper proposed a CNN accelerator based on a Multi-Mode Convolution Neural Network Unit (MMCN) in \cite{my}, Figure~\ref{fig:MMCN}. MMCN proposed a computation core that supports convolution, max-pooling, and dense functions. All these functions share the same hardware to increase the utilization of the circuit. However, there are two serious issues of MMCN:

\begin{itemize}
    \item Low efficiency on parallel CNN structure. The MMCN adopted a series strategy when facing a parallel CNN structure. Owing to operating each layer in series, MMCN takes much more time to complete the whole CNN model. However, more execution time results in more power consumption, which is against the target of a low-power CNN accelerator. 
    \item Large memory usage is caused by the large scale of input data in parallel CNN structure. There isn't any data reuse structure or technique in MMCN. As mentioned in \cite{ISSCC}, data transmission between core and memories has the most power of a chip. Therefore, although MMCN performs low computation power consumption, the total power of hardware still does not decrease apparently due to the huge requirement for memories.   
\end{itemize}

In \cite{ISSCC}, a dataflow inspired by \cite{pipe_resnet_1} and \cite{pipe_resnet_2} was proposed. This strategy re-arranges parallel structure to series, similar to the series method introduced before. However, it adds the pipeline technique between each layer. In a single operation cycle, \cite{ISSCC} can execute each convolution in the pipeline. Therefore, after conducting a parallel structure, take a residual block as an example, final outputs can be generated to output buffers. With this technique, \cite{ISSCC} speeds up the operation latency effectively. However, due to the modified series strategy, the data transmission of data and memories remained high. The proposed SF structure in this paper reduces the power of data transmission apparently, which will be introduced in detail in the next section.

For the diffusion model, the training flow is completed by adding noise to the original for a great scale of iteration while prediction is the opposite of training. The noise has to be removed from the previous feature map input to achieve the final clear output. This action is called "de-noise" in the diffusion model. The accelerator has to conduct thousands or even millions of times to get the output figure. To achieve this goal, power consumption and latency are the most difficult issues for traditional accelerators. Therefore, this paper proposed a structure that is equipped with the ability to conduct de-noise computation under fast, low-power figures. 

\begin{figure}[H]
\centering
\includegraphics[width=0.45\textwidth]{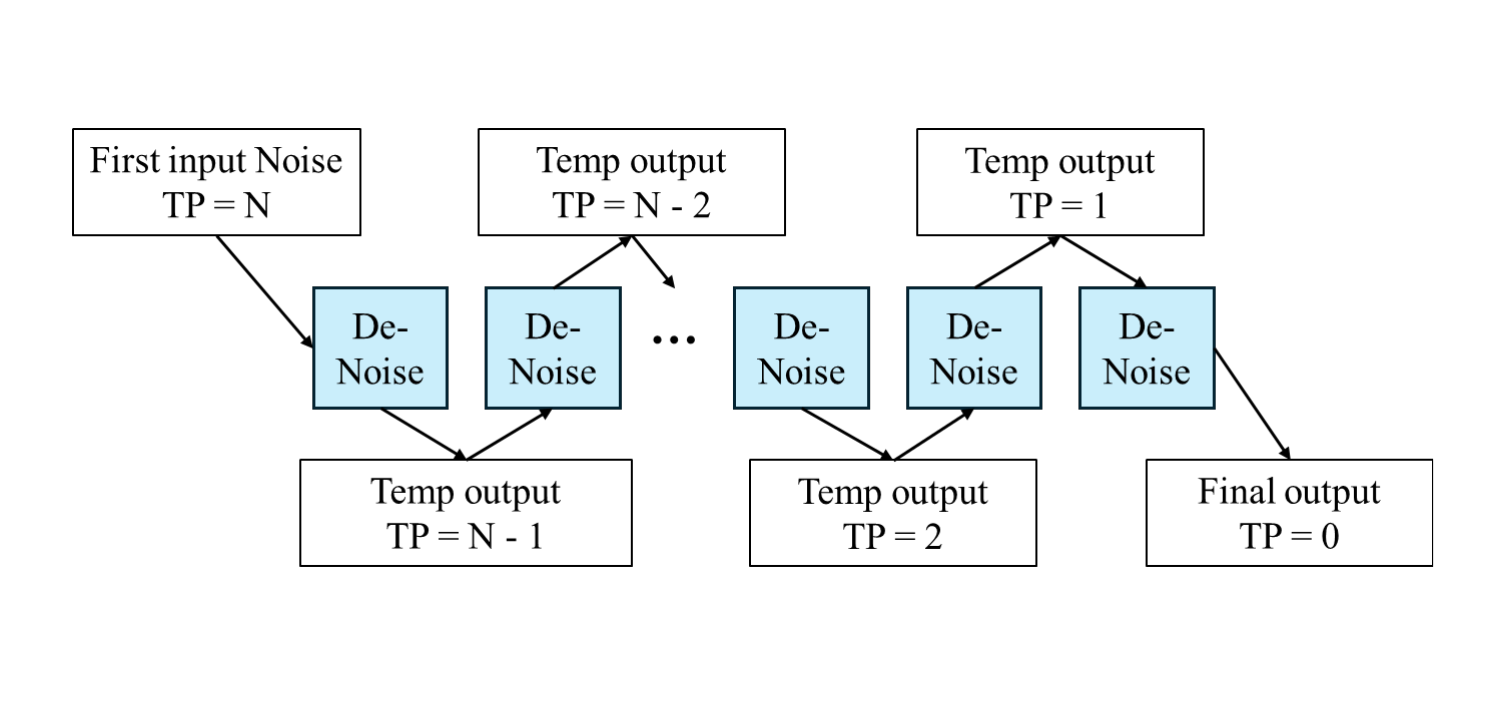}
\caption{De-noise dataflow of diffusion model \cite{diffusion}.} 
\label{fig:denoise}
\end{figure}

\section{Sever Flow Multi-Mode Convolution Neural Network Unit}
The design and implementation of SF-MMCN will be discussed in this section.

\subsection{PE design}
\begin{figure}[H]
\centering
\includegraphics[width=0.45\textwidth]{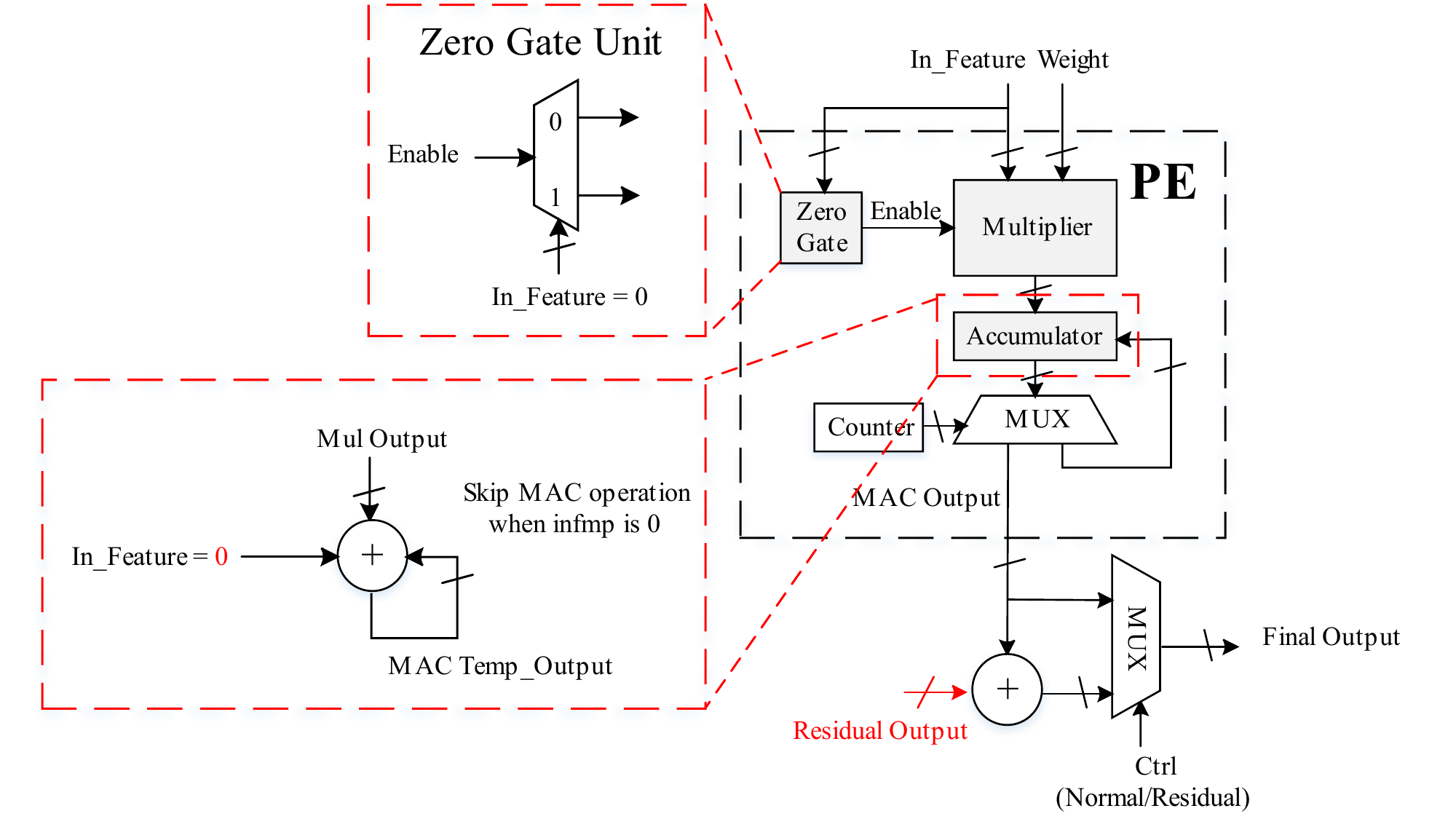}
\caption{A PE in the proposed-2 structure.} 
\label{fig:P2PE structure}
\end{figure}

A PE in SF-MMCN is shown in Figure~\ref{fig:P2PE structure}. The most apparent modification of a PE is the selection between normal convolution output and convolution output with residual block. If the SF-MMCN is under the residual mode, the MAC outputs will enter an adder to finish convolution computing with residual outputs. The MAC outputs will bypass to output registers to complete the normal convolution. A PE in SF-MMCN also equipped pipeline technique by a counter, which can enhance the throughput of whole structures. \par

The proposed PE is also equipped with a zero gate unit which can detect input image data. If input image data is zero, the zero gate unit will turn off a multiplier and skip MAC operation to avoid redundant energy consumption. With a zero gate unit, the proposed PE can reduce additional power dramatically. \par

\subsection{Pipeline Technique}
As depicted in Figure~\ref{fig:P2PE structure}, a pipeline technique is integrated into each PE within the proposed SF-MMCN structure. Diverging from conventional CNN accelerators, including those in PE arrays like \cite{ISCAS}, the SF-MMCN structure optimizes chip area by incorporating the pipeline technique into each PE. Unlike the traditional approach of passing MAC temporary outputs sequentially from one PE to the next, the MMCN structure efficiently bypasses these computation paths through the utilization of the pipeline technique. To enhance the overall utilization and configuration of the proposed SF-MMCN structure, this study strategically employs counters to facilitate the implementation of the pipeline function.


\begin{figure}[htbp]
\centering
\includegraphics[width=0.45\textwidth]{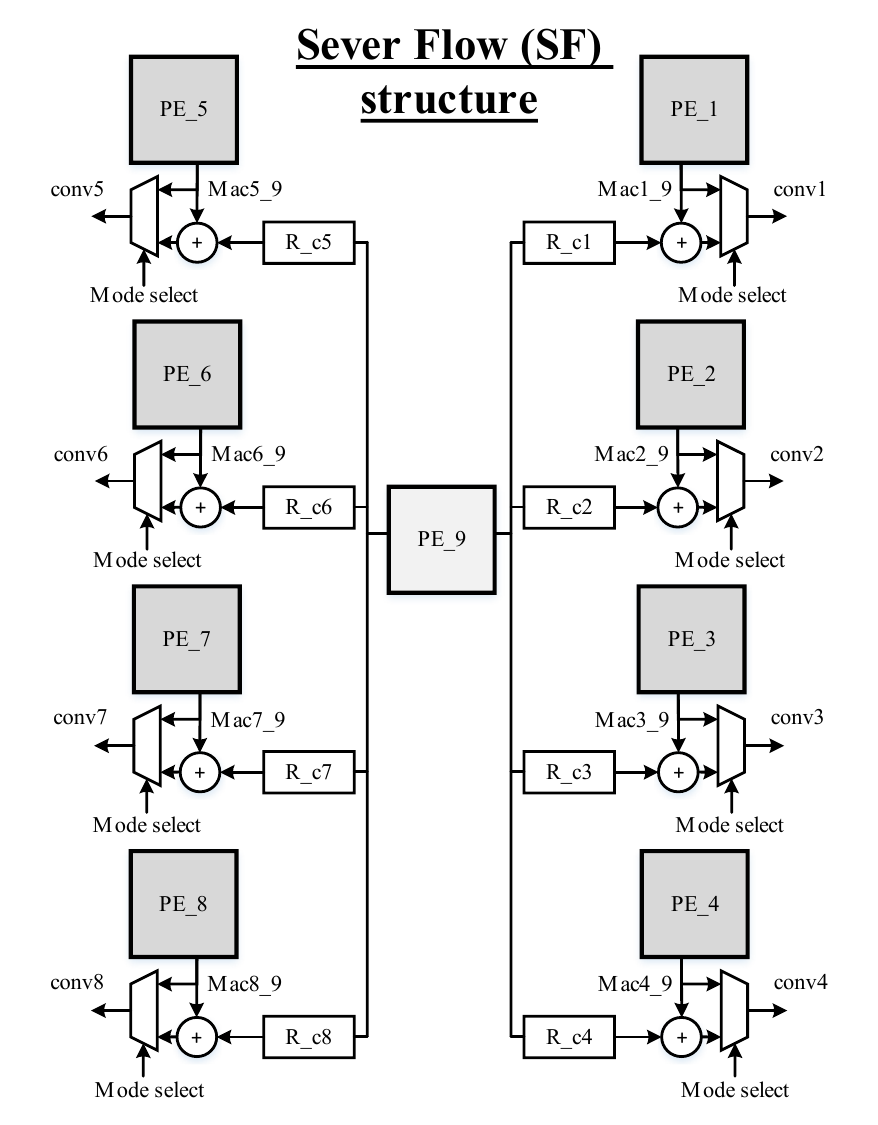}
\caption{Sever Flow (SF) structure.} 
\label{fig:SF structure}
\end{figure}

\begin{figure}[htbp]
\centering
\subfloat[]{
\includegraphics[width=0.4\textwidth]{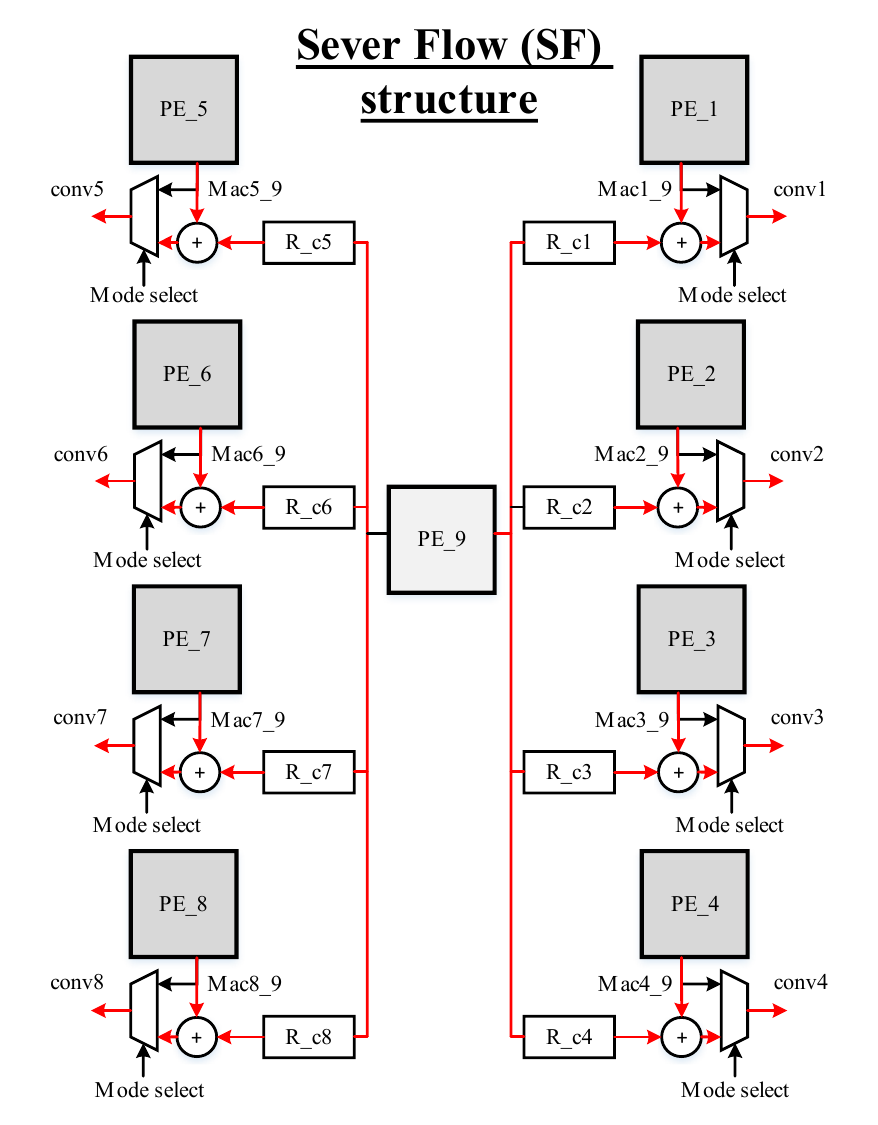}}
\hfill
\subfloat[]{
\includegraphics[width=0.35\textwidth]{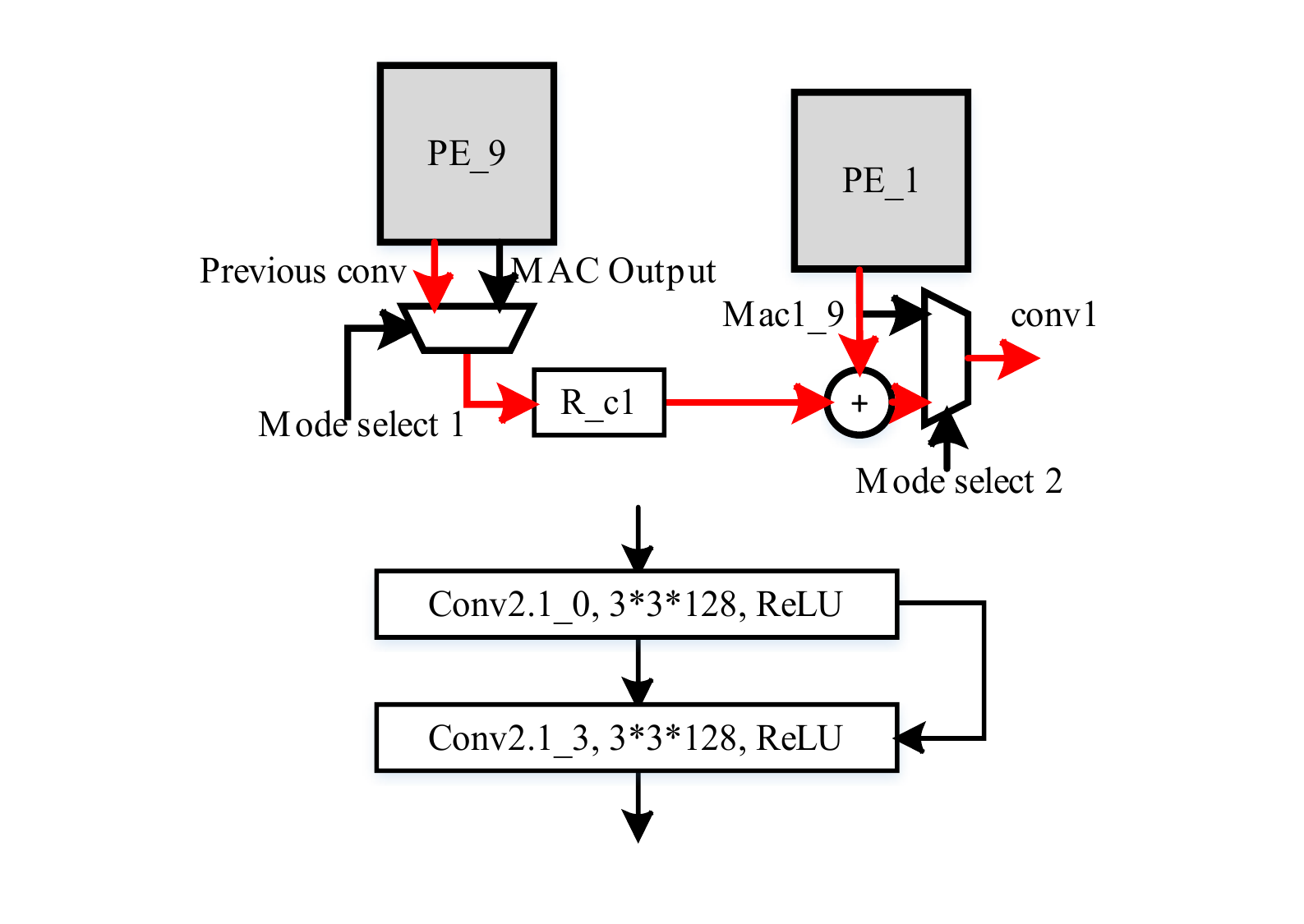}}
\\
\subfloat[]{
\includegraphics[width=0.35\textwidth]{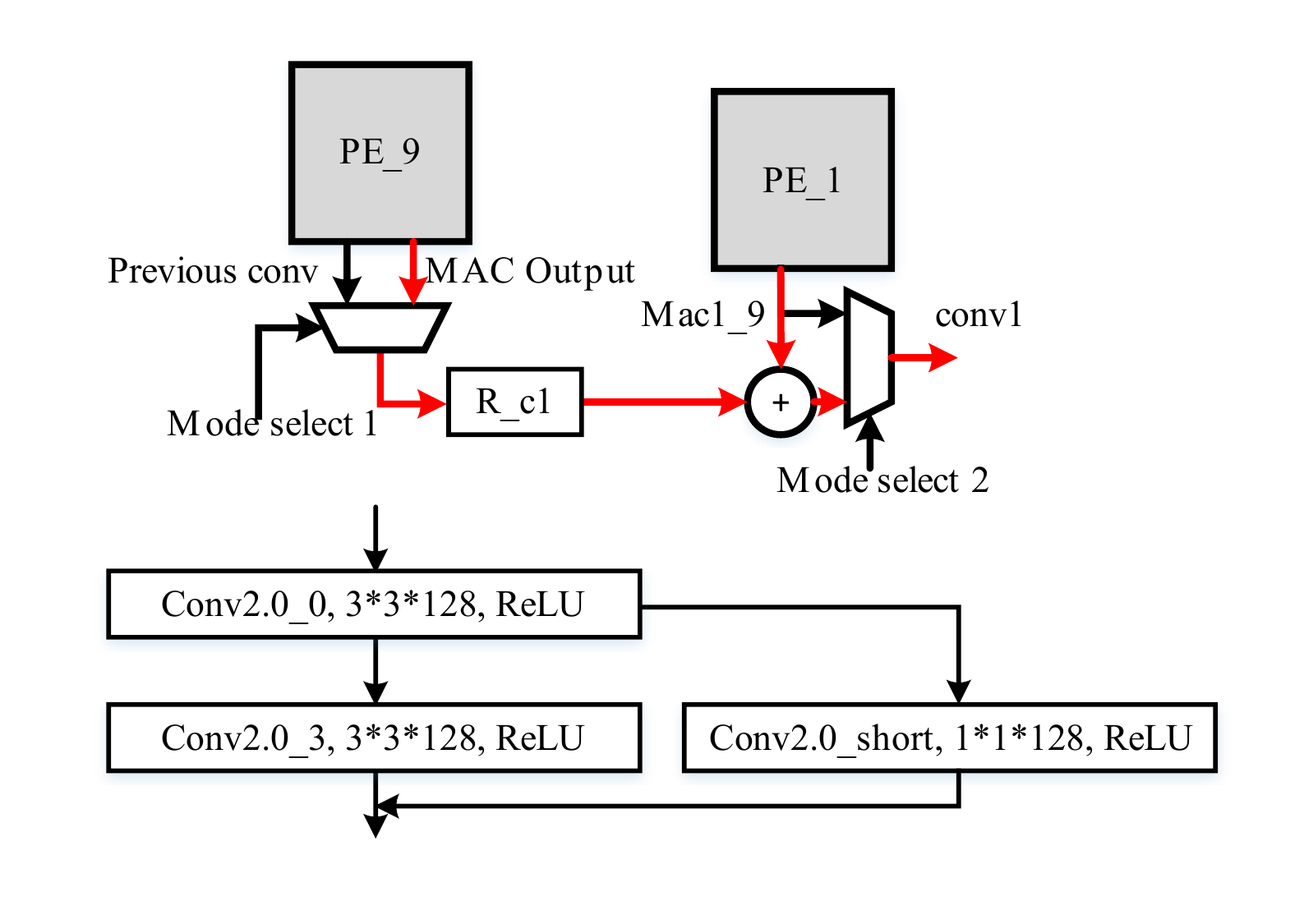}}
\caption{(a) The dataflow of SF while executing. (b) The closer look of SF structure under normal residual structure. (c) The closer look of SF structure under residual structure with a convolution layer.} \label{fig:SF_intro}
\end{figure}

\subsection{Sever Flow}

As previously discussed, within the PE array of the SF-MMCN architecture, one PE is assigned the responsibility of transmitting data to the other eight PEs, as illustrated in Figure~\ref{fig:SF structure}. The data flow during this process is demonstrated when a PE is passing data. During the computation of a series CNN model, the PE is typically set into idle mode to minimize power consumption. However, when confronted with a parallel CNN structure, SF initiates the corresponding computations, as depicted in Figure~\ref{fig:SF_intro} (a) and Figure~\ref{fig:SF_intro} (b).\par

The primary challenge in SF-MMCN lies in concurrently achieving normal convolution and a residual block without introducing any additional circuitry. Traditional CNN accelerators often compromise either circuit area or latency to execute parallel convolution. Consequently, the PE array structure in SF-MMCN deviates from conventional designs, adopting a server-flow (SF) configuration. As exemplified in Figure~\ref{fig:SF structure}, PE\_9 serves as a server, delivering the corresponding output of a residual block from PE\_1 to PE\_8. This unique configuration enables SF-MMCN to seamlessly integrate both normal convolution and residual block functionalities without the need for supplementary circuitry.\par

During normal convolution operations in SF-MMCN, PE\_9 is deactivated. The datapath for this scenario, along with the corresponding CNN structure, is illustrated in Figure~\ref{fig:SF_intro} (a). The MAC output, denoted as MAC1\_9, swiftly enters a multiplexer, depicted by the red line in Figure~\ref{fig:SF_intro} (a), to seamlessly complete the normal convolution process. While PE\_9 may be considered redundant in this particular function, its deactivation contributes to the overall efficiency of the PE array, allowing for high-efficiency multi-mode operation through selective mode choices.


During the execution of the residual function in SF-MMCN, without the presence of a residual convolution block, PE\_9 becomes active. The hardware and software datapath for this scenario is depicted in Figure~\ref{fig:SF_intro} (b) an d (c). In this operation, designed for residual function without a residual convolution block, Figure~\ref{fig:SF_intro} (b), the objective is to accumulate previous convolution outputs and the MAC outputs from PE\_1 to PE\_8. Consequently, PE\_9 is specifically tasked with transmitting the previous convolution output to a multiplexer controlled by the testbench (Mode select 1). The output from PE\_9 then enters an adder situated near the rest of the PEs, where it accumulates with a MAC output (MAC1\_9) to complete the computation. This configuration streamlines the process, facilitating the efficient execution of residual functions within the SF-MMCN architecture.

\begin{figure}[H]
\centering
\includegraphics[width=0.48\textwidth]{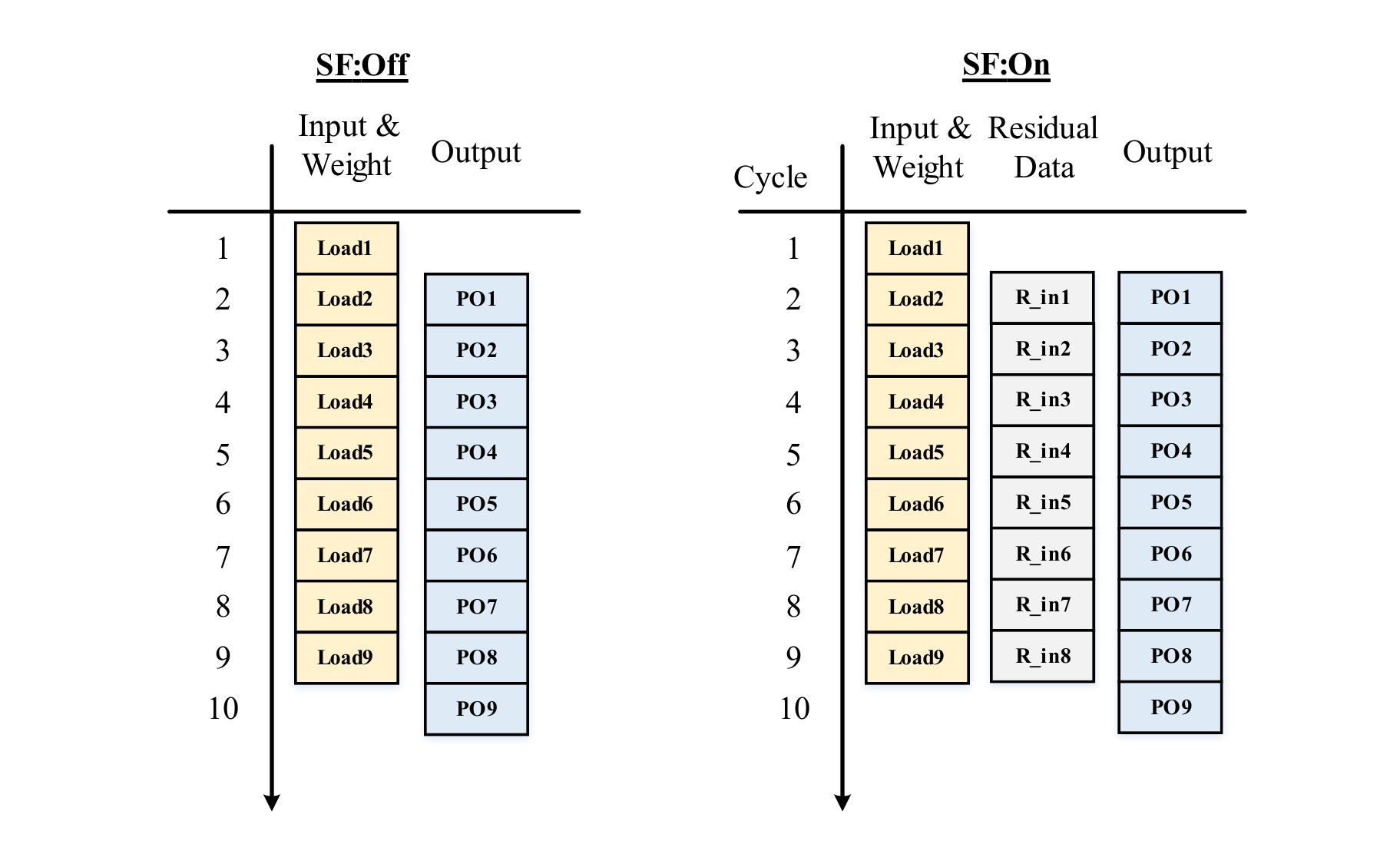}
\caption{The dataflow of the proposed SF on/off.} 
\label{fig:SFwave}
\end{figure}


When SF-MMCN is engaged in the residual function with a residual convolution block, the output path of PE\_9 transitions to its dedicated Multiply-Accumulate (MAC) output, as illustrated in Figure~\ref{fig:SF_intro} (c). This output, along with the MAC outputs from PE\_1 to PE\_8, is directed to an adder to accomplish convolution. In this capacity, PE\_9 functions akin to a server, tasked with preparing the residual convolution output. Importantly, considering the shape of weights in residual convolution, SF can efficiently set up the preparation of residual convolution output promptly, ensuring synchronization with the completion of MAC computations by the other eight PEs.

SF-MMCN's ability to execute parallel CNN structures without requiring additional computation cycles results in notable advantages. It not only saves on memory access and data transfer between SF-MMCN and memory but also enables the simultaneous completion of normal convolution and residual functions with convolution. This multi-functionality is achieved while maintaining low power consumption, underscoring the efficiency and versatility of the SF-MMCN architecture.

\begin{figure*}[htbp]
\centering
\includegraphics[width=18cm]{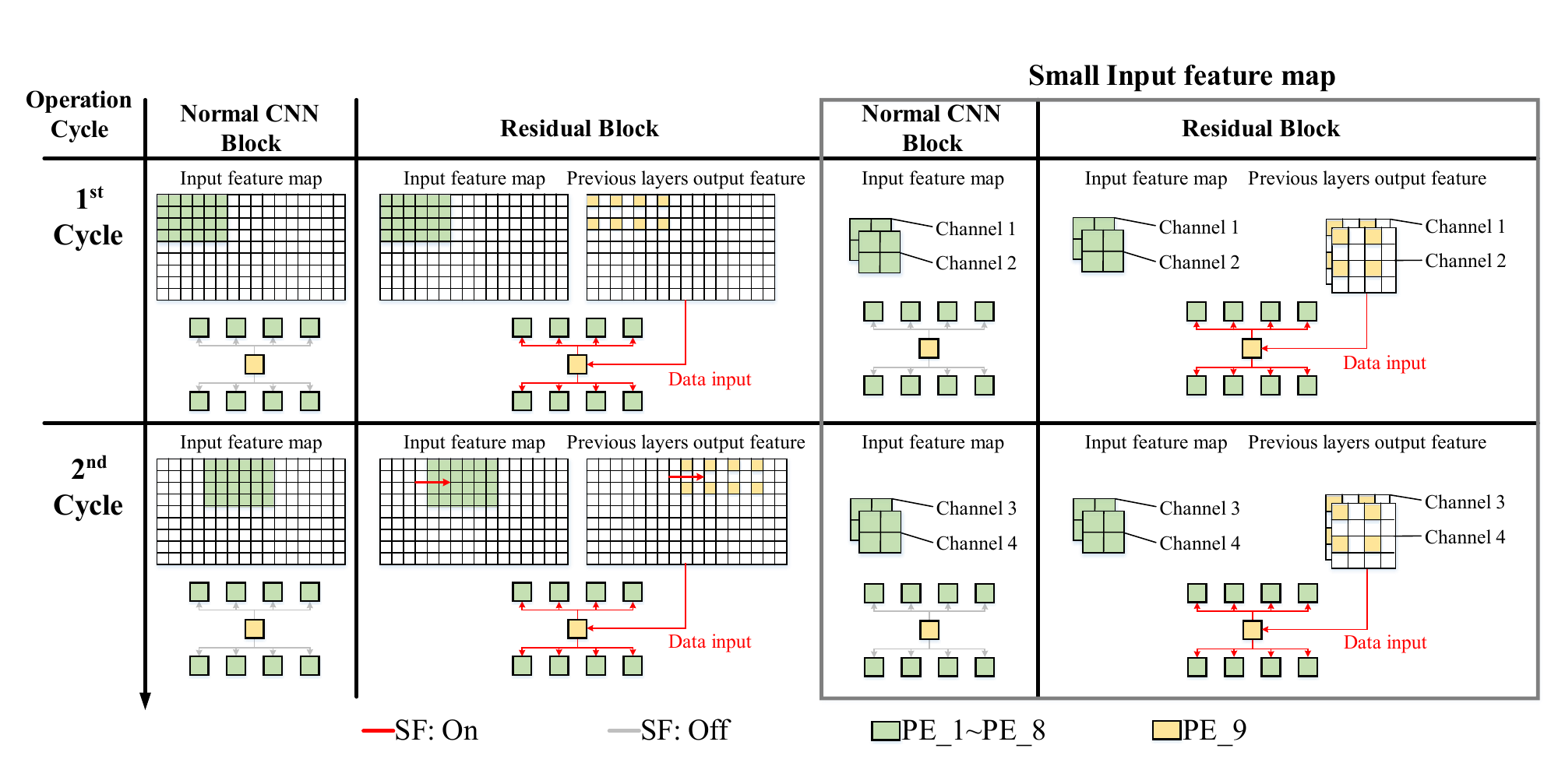}
\caption{The illustration of SF-MMCN with different size of input feature map.} 
\label{fig:SF-MMCN flow}
\end{figure*}

\subsection{Operation Flow}

Figure~\ref{fig:SFwave} shows the waveform of SF-MMCN, PO represents partial output. Single convolution can be completed in 10 cycles if the width and height of a filter are $3*3$. Once input and weight data enter SF-MMCNs, the MAC operation executes immediately. After 9 input features and weights finish loading, the proposed SF-MMCN will take 1 cycle to get the final convolution outputs. When encountering residual structure, the proposed SF-MMCN consumes the same cycles as normal convolution. As mentioned above, PE\_1 to PE\_8 compute all MAC data of normal convolution, and PE\_9 would deliver or compute the previous convolution output and MAC operation respectively during the same cycles. Therefore, the proposed SF-MMCN can complete the residual structure without redundant computation cycles.\par

\begin{figure}[htbp]
\centering
\includegraphics[width=0.48\textwidth]{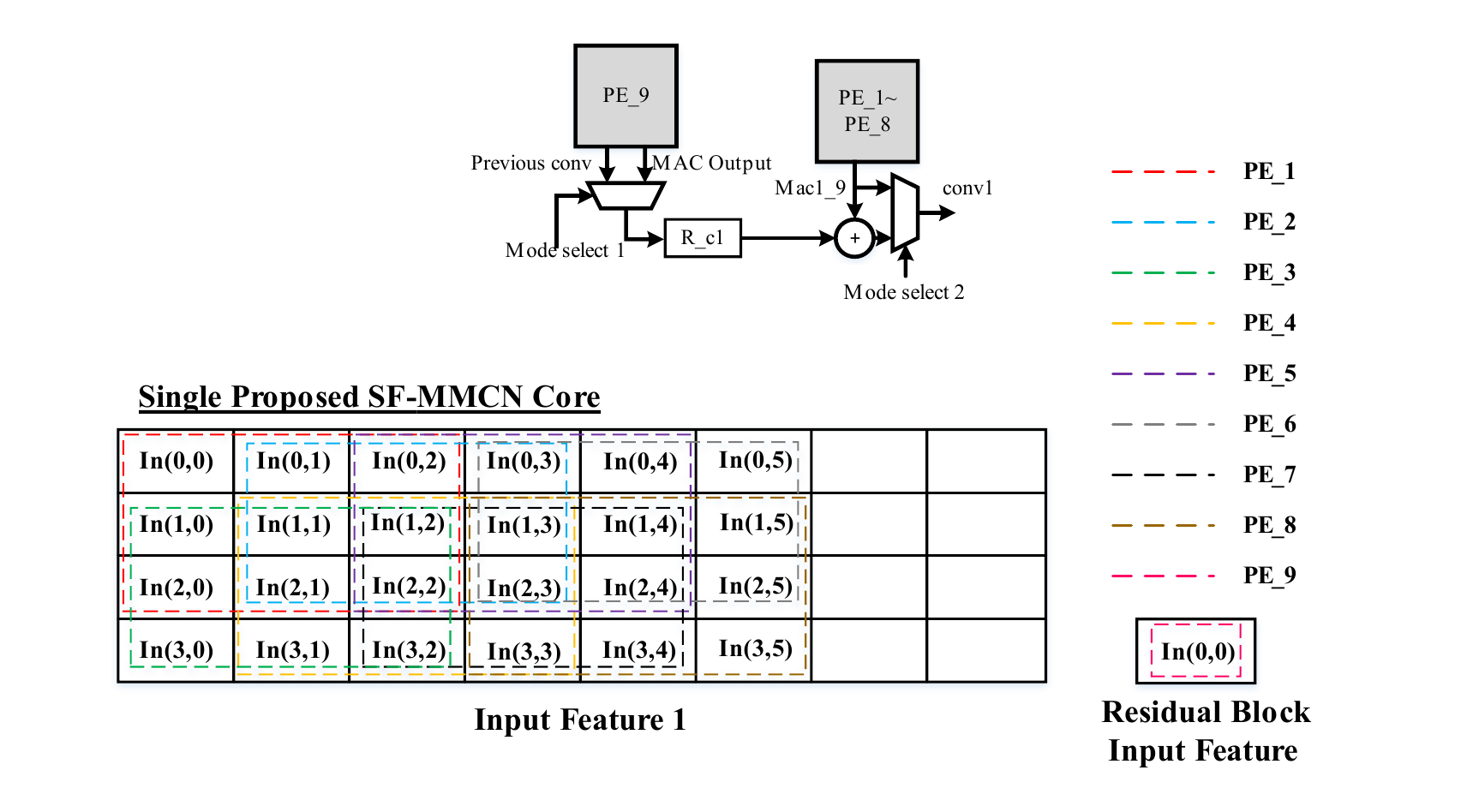}
\caption{Illustration of the input feature and corresponding executed PEs in the proposed SF-MMCN.} 
\label{fig:SFoutShpae}
\end{figure}

The proposed SF-MMCN performs high throughput due to the SF structure. Owing to four SF-MMCNs in implementation architecture Figure~\ref{fig:SF-MMCN}, the shape of output data by PE\_1 to PE\_8 is $3\times3\times8$, which means $width\times height\times channel$ and the output shape of PE\_9 is $1\times1\times8$, Figure~\ref{fig:SFoutShpae}. The value of the channel equals the number of the SF-MMCN in the implementation. The batch size of the proposed SF-MMCN is 1 because of the high-speed convolution requirement of CNN accelerators~\cite{CARLA}. In other words, due to the batch size being 1, the proposed SF-MMCN can provide decision results in real-time.

\begin{figure}[htbp]
\centering
\includegraphics[width=0.48\textwidth]{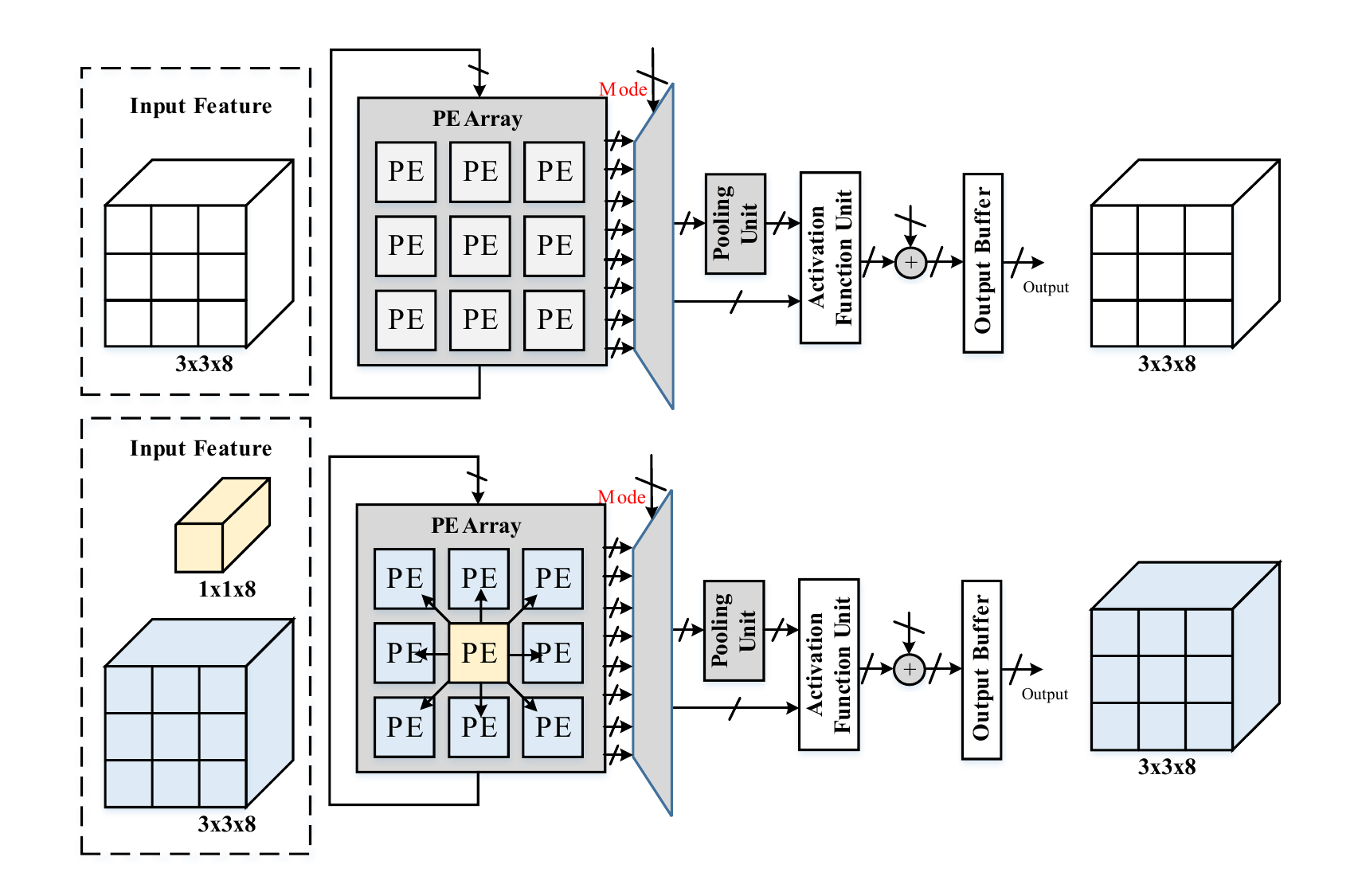}
\caption{The shape of input features of the proposed SF-MMCN under different conditions.} 
\label{fig:diffmode}
\end{figure}

As introduced above, Figure~\ref{fig:diffmode} shows the input data shape of the proposed under different environments. When encountering a series structure, the shape of the output is the same as the input shape with zero padding, $3*3*8$. While facing parallel CNN structures, the throughput increases because PE\_9 executes the input feature of the parallel block. The allocation of an input feature map is indicated in Figure~\ref{fig:diffmode} with different colors.\par

\begin{figure}[H]
\centering
\includegraphics[width=0.45\textwidth]{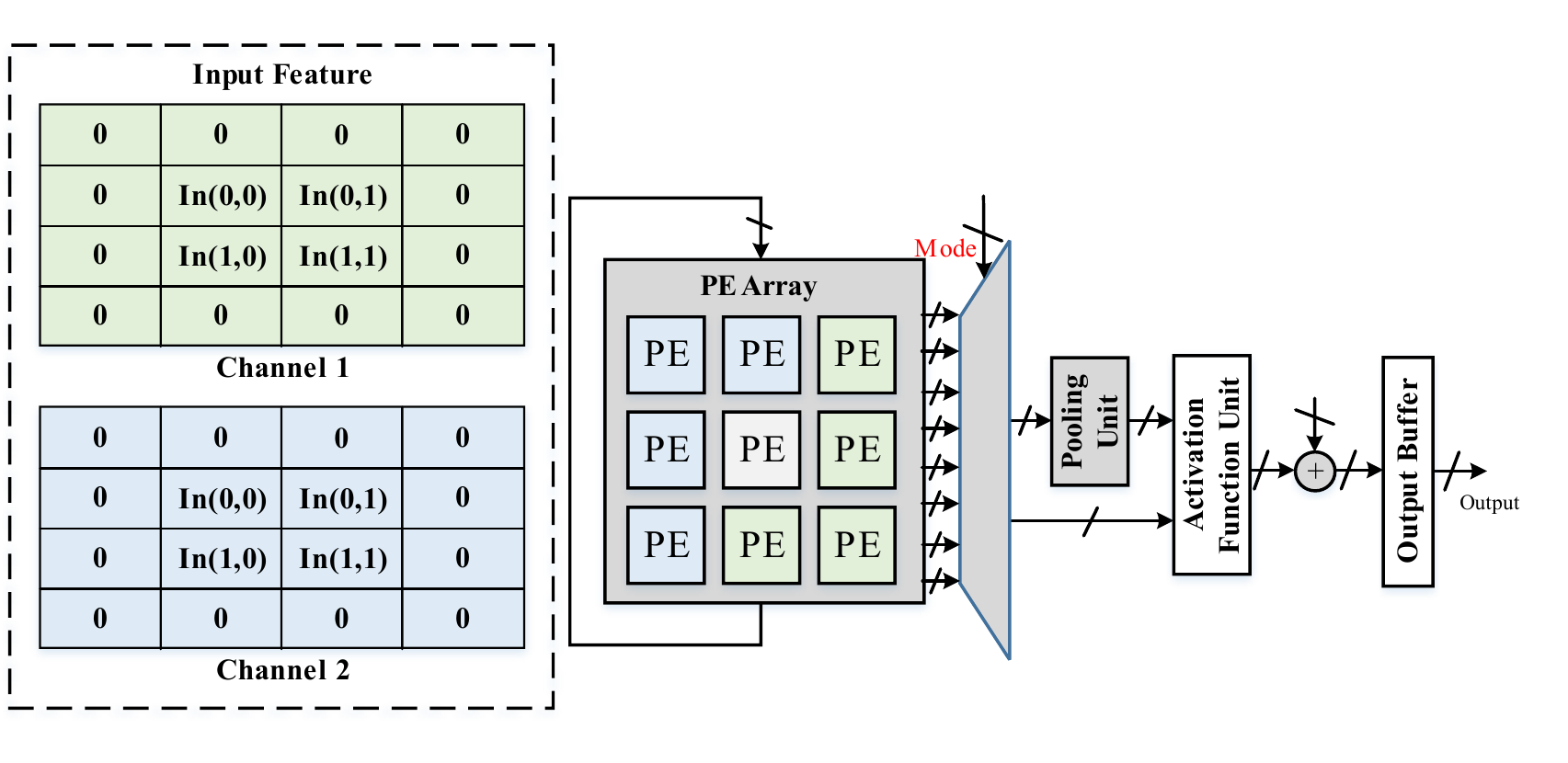}
\caption{The dataflow of the proposed SF-MMCN on a small size input feature.} 
\label{fig:smallmap}
\end{figure}

Another problem is that the proposed can address the input features with a small size. Take $2\times2$ input map as an instance, the PE array will separate into two parts to complete two channels of input data, Figure~\ref{fig:smallmap}. Therefore, the proposed SF-MMCN reduces the potential of redundant circuits under different complexities of the environment.

Consequently, Figure~\ref{fig:SF-MMCN flow} demonstrates the datapath of the proposed SF-MMCN with a normal and small input feature map under both common CNN structures and residual blocks. The green and yellow squares are responsible for series and parallel structure respectively.

As the address of an input feature map of the residual block is not continued, the address can be generated by shifters. The corresponding input data pass into the PE\_9 directly to finish MAC computing. According to Figure~\ref{fig:SF-MMCN flow}, it's straightforward forward in implement eight input data of a residual block that can be well-distributed to the rest of PEs to complete MAC operation. Therefore, SF-MMCN doesn't waste any additional cycle waiting for MAC output from PE\_1 to PE\_8. \par

As described in Figure~\ref{fig:smallmap}, the control unit separates the PE array into two parts to complete computations. When facing residual structure, the PE\_9 of the proposed SF-MMCN can complete two channels of computations before the MAC outputs are delivered from other PEs due to the same pixel number of both series block and parallel block.

\begin{figure}[H]
\centering
\includegraphics[width=0.45\textwidth]{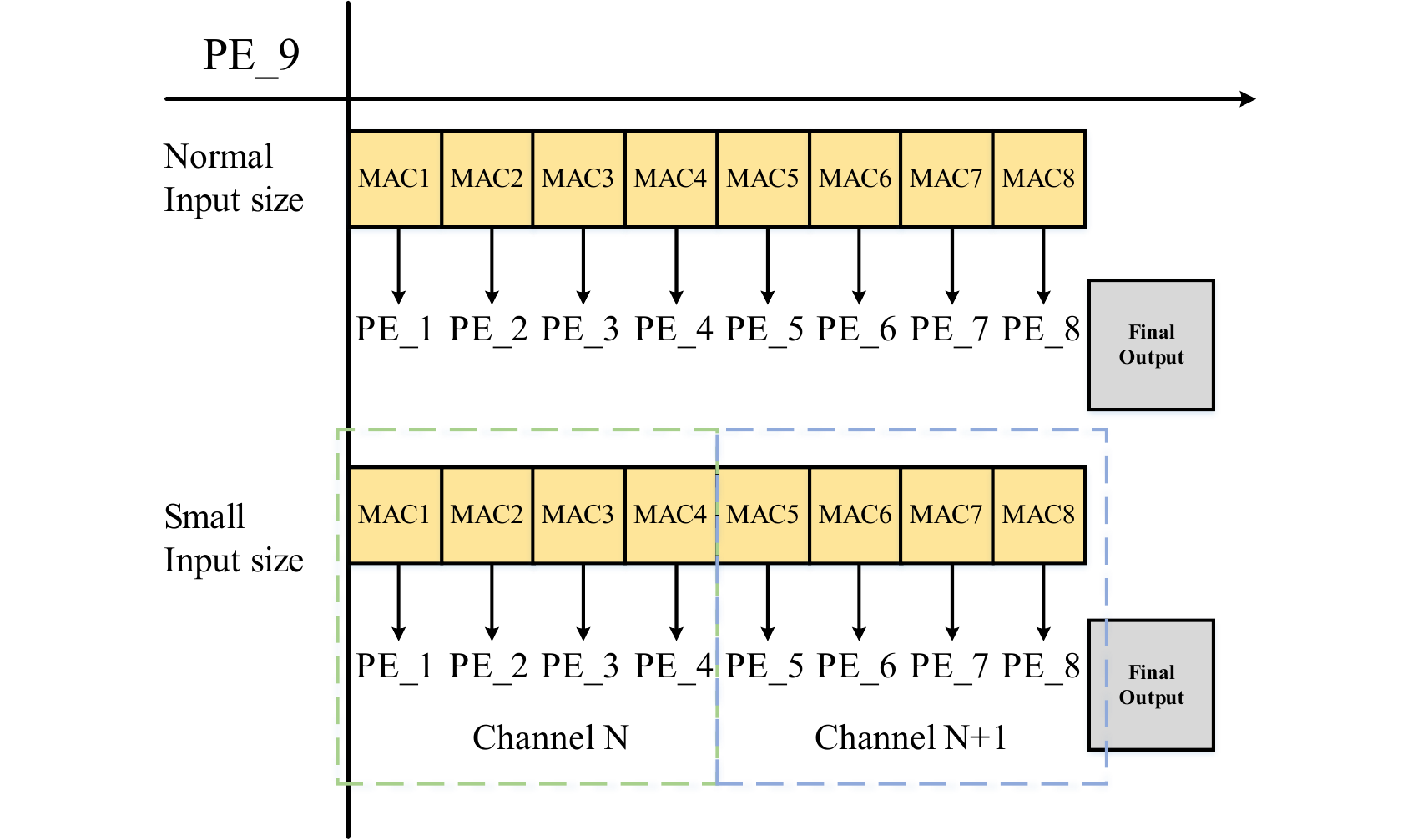}
\caption{The dataflow of PE\_9 on a normal/small size input feature.} 
\label{fig:normal/smallmap}
\end{figure}

Figure~\ref{fig:normal/smallmap} indicated the dataflow of PE\_9 under the different sizes of input. Before the whole system receives the final answer, PE\_9 passes the corresponding MAC data to each PE under normal input image size conditions. On the other hand, when the condition changes to a small input image size, channel N and channel N+1 in Figure~\ref{fig:normal/smallmap} is the distribution separated by the control unit. At the beginning of four cycles, PE\_9 is responsible for the computations of channel N, the rest of the four cycles are the computations of the MAC data of channel N+1. Therefore, due to the hardware and timing distribution of the proposed SF-MMCN, there aren't any occurrences of redundant circuits and cycles. 

\subsection{Diffusion model}
\begin{figure}[H]
\centering
\includegraphics[width=0.45\textwidth]{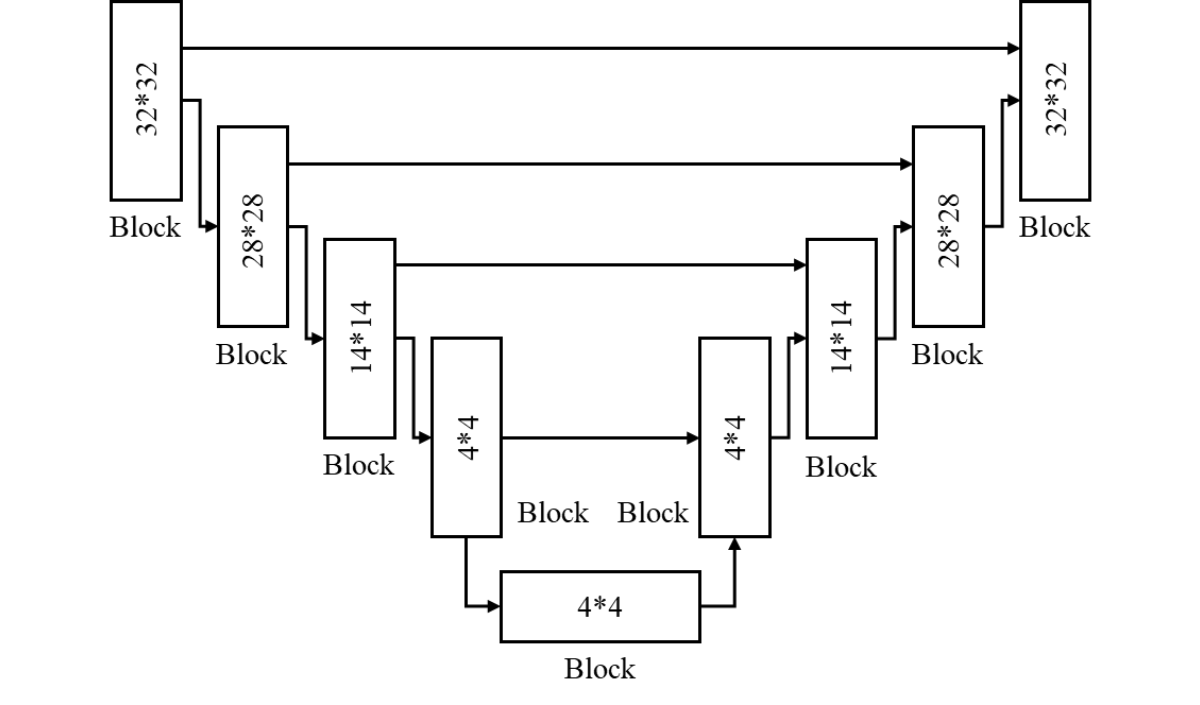}
\caption{The U-net structure.} 
\label{fig:punet}
\end{figure}

\begin{figure}[H]
\centering
\includegraphics[width=0.45\textwidth]{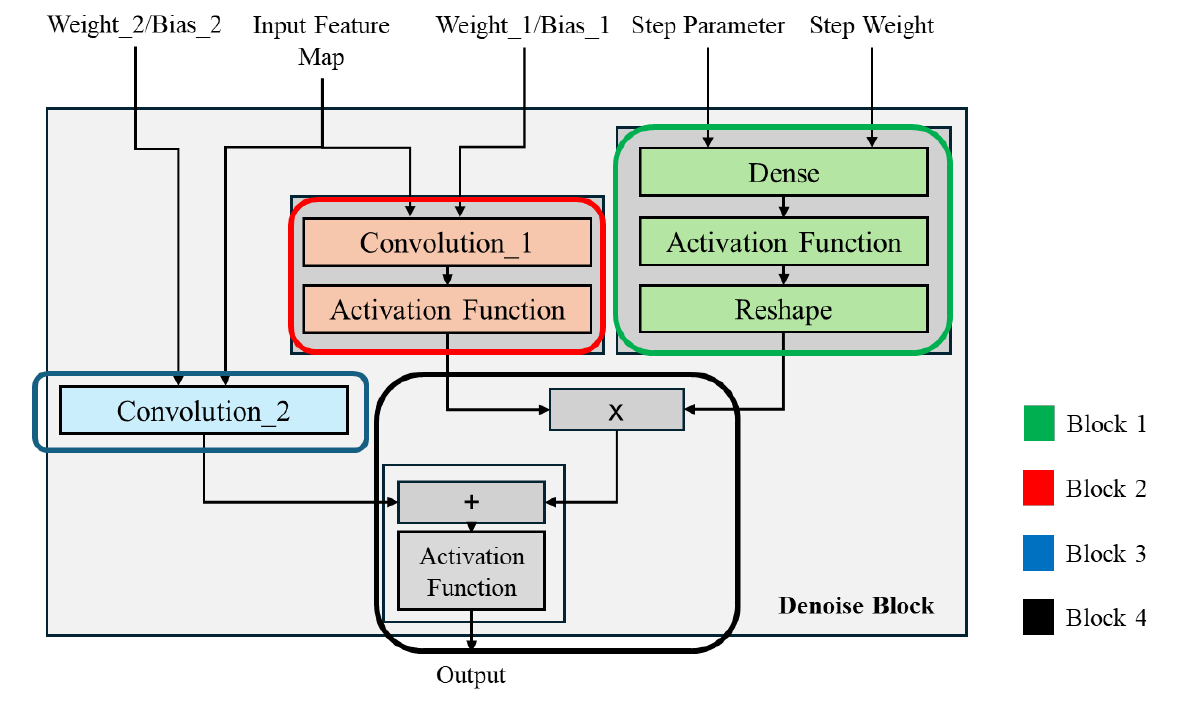}
\caption{The block distribution of U-net.} 
\label{fig:blk_structure}
\end{figure}

Figure~\ref{fig:punet} shows the U-net which is implemented in this paper. There are lots of blocks in the model. Each block consists of two convolution layers and one dense layer which is responsible for the computation of time parameters. Moreover, the hardware of each block is the same, which means that the proposed SF-MMCN can support multiple computations in the same operation cycle. The dataflow of a block of U-net is indicated in Figure~\ref{fig:blk_structure}. There are 4 groups of a single block, which distributes the U-net block into a dense layer (Block 1), a convolution layer with an activation function (Block 2), a convolution layer without an activation function (Block 3), and final logic computation (Block 4).  

\begin{figure}[H]
\centering
\includegraphics[width=0.45\textwidth]{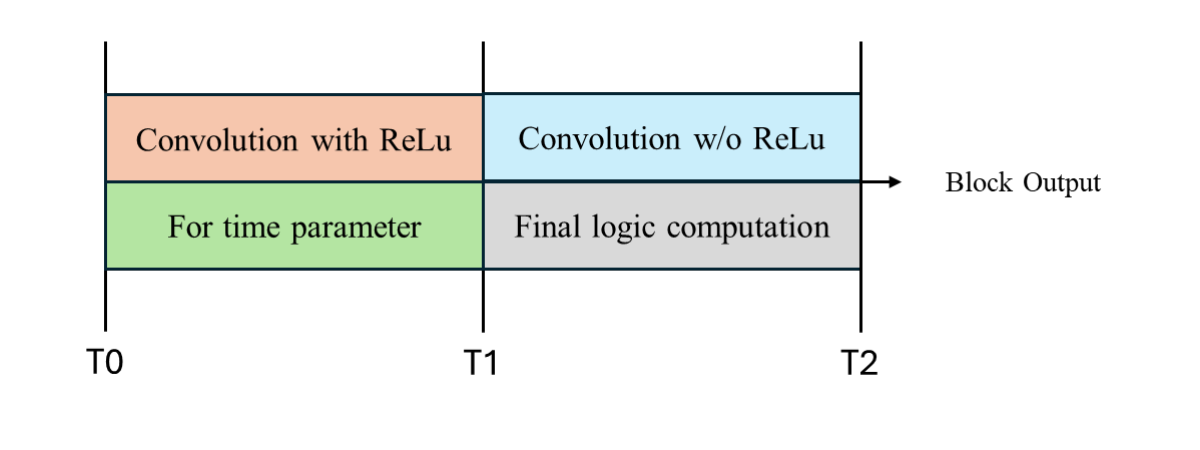}
\caption{The operation timing distribution of the proposed SF-MMCN.} 
\label{fig:time}
\end{figure}

\begin{figure}[H]
\centering
\includegraphics[width=0.45\textwidth]{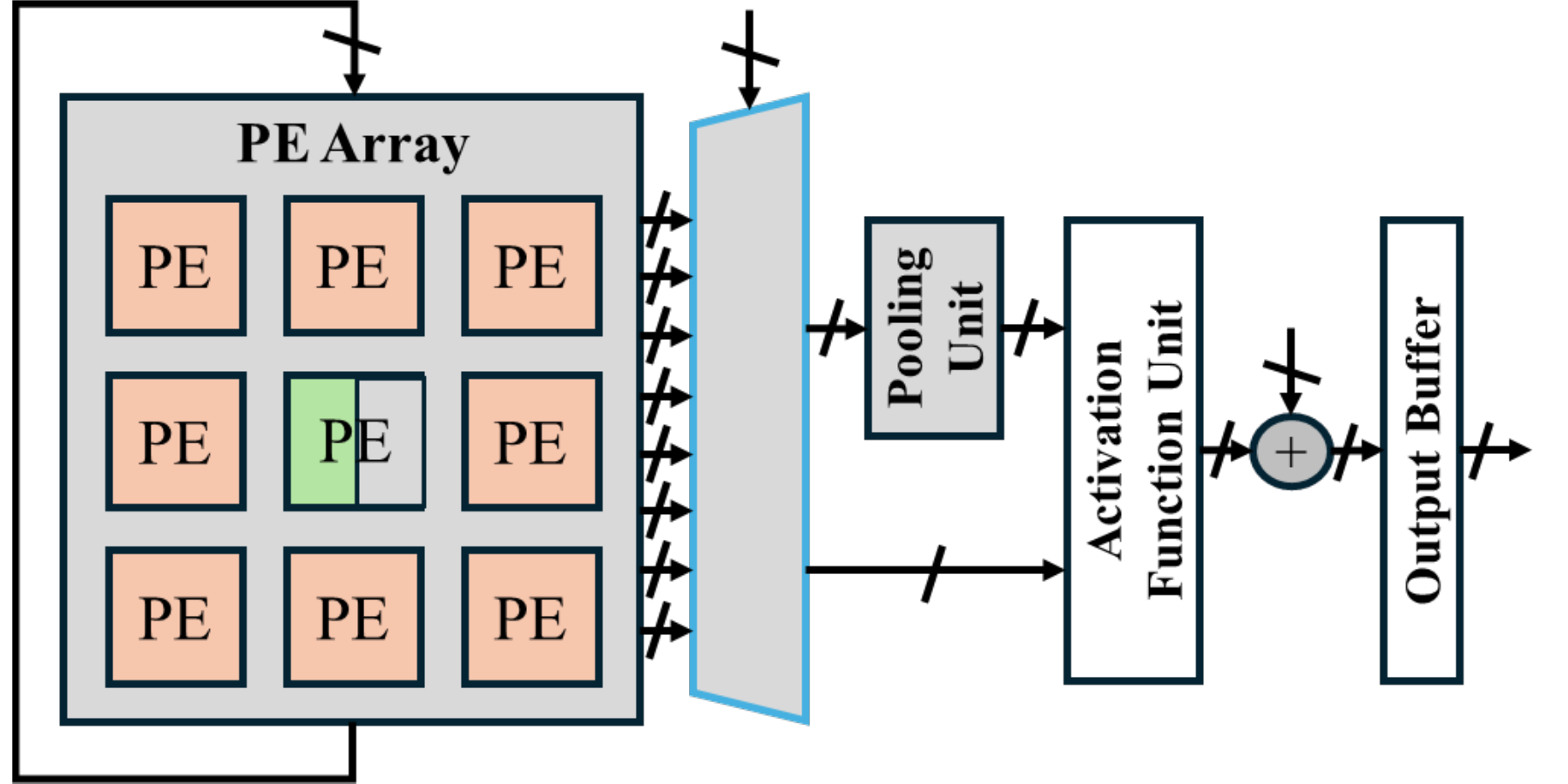}
\caption{The allocation of the proposed structure among different blocks.} 
\label{fig:sf_diff}
\end{figure}

According to the block distribution, the dataflow with timing is described by Figure~\ref{fig:time} and Figure~\ref{fig:sf_diff}. When the computation starts (T0 in Figure~\ref{fig:time}), Figure~\ref{fig:sf_diff} PE\_1 to PE\_8 (orange) are responsible for the convolution with activation function (ReLu). Meanwhile, PE\_9 conducts a time parameter dense layer. From T1 to T2 in Figure~\ref{fig:time}, PE\_1 ~ PE\_8 operates the other convolution layer and combines the final result through final logic computation.

By introducing the implementations of the proposed SF-MMCN, the key to the proposed SF is PE\_9. Since the rest of the PEs are recommended to fulfill the main connection of models, PE\_9 is assisted to deal with the parallel structure such as residual and time parameter operator. This architecture seems not quite complex but can complete difficult, deep learning models. Therefore, the proposed structure is equipped with great configurable features.

\subsection{Data Reuse}
\begin{figure}[htbp]
\centering
\subfloat[]{
\includegraphics[width=0.48\textwidth]{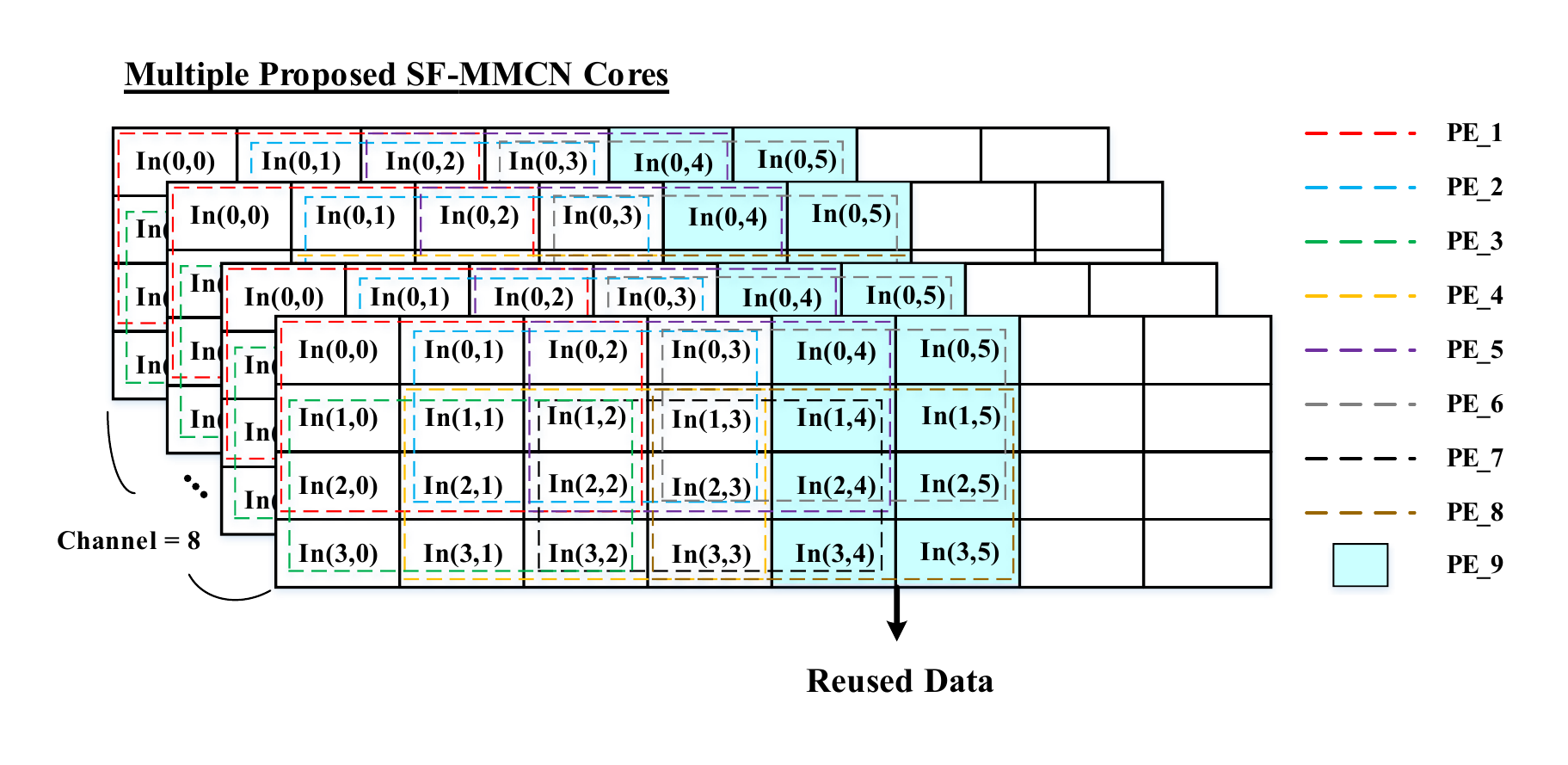}}
\hfill
\subfloat[]{
\includegraphics[width=0.45\textwidth]{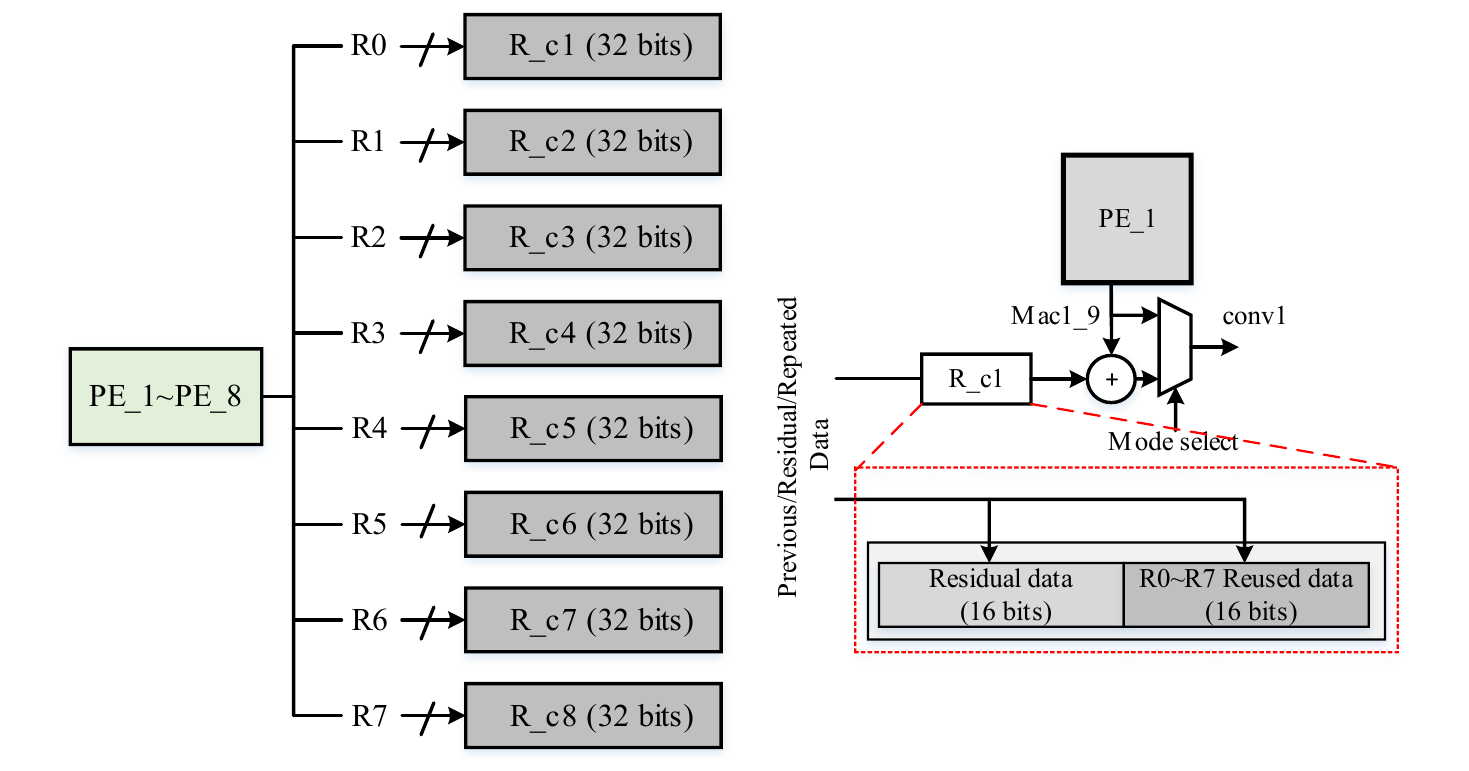}}
\caption{(a) Illustration of repeated data in the proposed SF-MMCN. (b) Data-reused structure.} \label{fig:datareuse}
\end{figure}

Leveraging the SF structure, the proposed SF-MMCN optimizes data reuse due to the prevalence of repeated input images, Figure~\ref{fig:datareuse} (a). Between each convolution cycle, there exist 8 repeated input data, corresponding to 8 registers from PE\_1 to PE\_8. To accommodate the 16-bit requirements for residual blocks in hardware, SF-MMCN expands the bit-width of these 8 registers to 32 bits, facilitating simultaneous storage of residual and reused data, Figure~\ref{fig:datareuse} (b). This strategic approach reduces power consumption by avoiding the reloading of repeated data in every computation cycle. As a result, SF-MMCN significantly conserves energy, demonstrating the efficacy of its SF structure.

\subsection{Implementation Architecture}

\begin{figure*}[htbp]
\centering
\includegraphics[width=14cm]{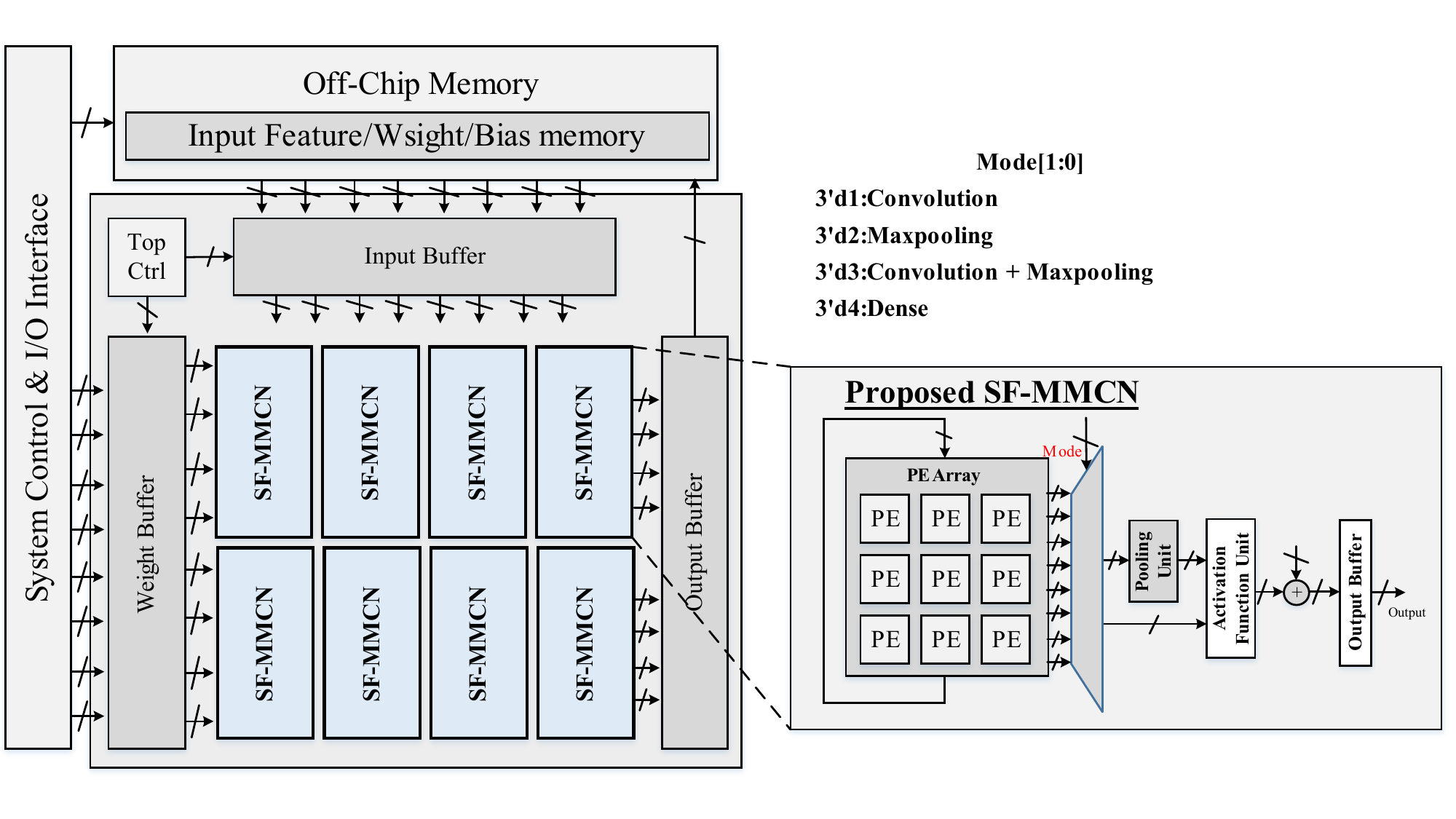}
\caption{The architecture of SF-MMCN.} 
\label{fig:SF-MMCN}
\end{figure*}

The implementation of the proposed SF-MMCN is indicated in Figure~\ref{fig:SF-MMCN}. The input image features and weight data are stored in off-chip memory before executing. After input and weight data enter the computation core through I/O interface, dataflow would be managed by TOP CTRL through weight and input buffer.  \par

\begin{figure}[htbp]
\centering
\subfloat[]{
\includegraphics[width=0.45\textwidth]{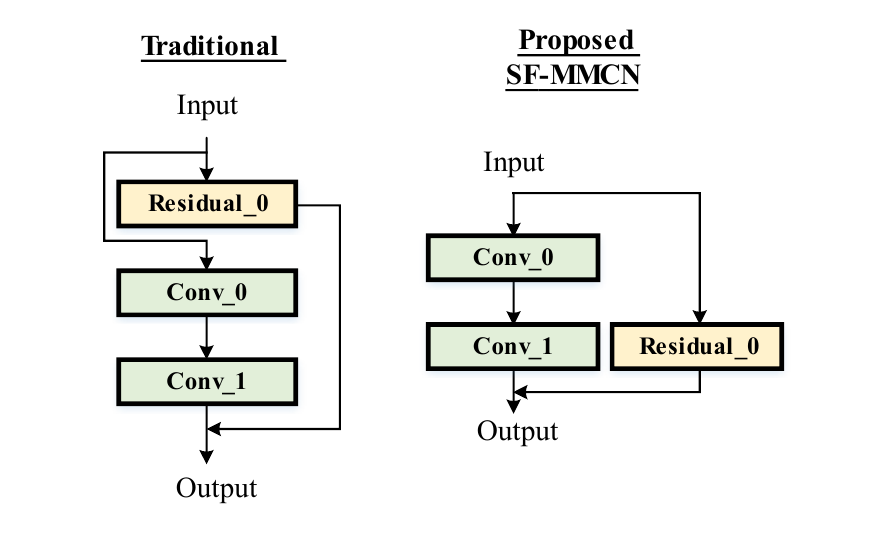}}
\\
\subfloat[]{
\includegraphics[width=0.45\textwidth]{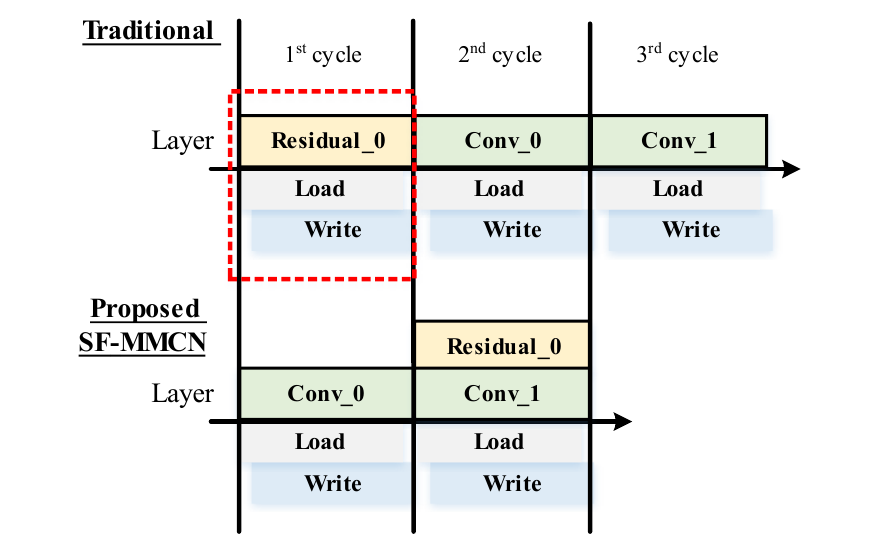}}
\caption{Dataflow comparison between traditional structures and the proposed SF-MMCN on (a) waveform. (b) CNN structure illustration.} 
\label{fig:wavecom}
\end{figure}

There are four SF-MMCNs in the implementation of the proposed SF-MMCN. Owing to data reuse in implementation and large MAC in a dense layer, each SF-MMCN can exchange data by registers of each PE in SF-MMCN. If the output channel exceeds hardware limitation, the dataflow of computation will iterate to complete convolutions.
What's more, with a pooling unit and activation function unit in the proposed SF-MMCN, it also supports different functions of CNN as MMCN in \cite{my}. All control signals are switched by TOP CTRL through the testbench.

\subsection{Operation Efficiency of Residual Blocks}

In traditional CNN accelerators, when facing a parallel CNN structure, most of the strategy is modifying the dataflow from parallel to series \cite{ISSCC}, \cite{my}, as shown in Figure~\ref{fig:wavecom} (a). In this study, the proposed SF-MMCN combines a normal convolution layer and a convolution layer hidden in the residual block. The Residual\_0 in figure~\ref{fig:wavecom} (a) doesn't have to wait for series structure (Conv\_0 and Conv\_1 in Figure~\ref{fig:wavecom} (a)) finishing computing. In other words, according to Figure~\ref{fig:wavecom} (b), traditional CNN accelerators spend additional cycles to complete convolution layers with parallel structures. Therefore, the proposed SF-MMCN performs the higher efficiency while conducting parallel structures.

\subsection{Hardware Utilization and Power Efficiency}

This paper aims to increases the utilization of PEs in PE array in a CNN accelerator. Due to the pipeline technique in a PE in the proposed SF-MMCN, a single PE can complete a convolution computation by itself. The utilization of PE in hardware is indicated in: 

\begin{equation}
    \label{Ct}
    C_t = \frac{T}{t} \times 100\%
\end{equation}
\begin{equation}
    \label{UPE}
    U_{PE} = \frac{PE_{act}}{PE_{total}} \times C_t \times 100\%
\end{equation}

In equation~(\ref{Ct}), Where $C_t$ is the percentage of computing cycles, which means actual operating cycles $T$ divided by total enable cycles $t$. $PE_{act}$ and $PE_{total}$ represent actual executing PEs and total PEs in hardware respectively. $U_{PE}$ is the utilization of PEs in hardware. Although most CNN accelerators perform high PE utilization, the occurrence of redundant PE still happens when encountering an unsuitable CNN model.
The utilization of PE is relatively connected with power consumption. As the biggest structure in a CNN accelerator, the PE array consumes the most power in operation. The equation~(\ref{P_total}) indicates the total power $P_{total}$ of a CNN accelerator.

\begin{equation}
    \label{P_total}
    P_{total} = (N \times P_1) + P_R + P_C
\end{equation}

N is equal to $PE_{act}$ in equation~(\ref{UPE}), which means the number of actual executing PEs in hardware. $P_1$ represents the power consumption of a single PE, $P_R$, and $P_C$ are the power of redundant circuit and control unit respectively. However, given the diversity of conditions in different CNN models, such as the parallel block mentioned before. What's more, with the development of CC accelerators, many more studies perform high efficiency in implementation. $P_{total}$ only expresses the estimation of total power. Therefore, another concept of efficiency factor ($\nu$) is explained in equation~(\ref{NU}).

\begin{equation}
    \label{NU}
    Efficiency\;Factor (\nu) = \frac{P_{total}}{U_{PE}}
\end{equation}

The efficiency factor equals the utilization of PE divided by total power consumption. According to equation~(\ref{NU}), it reveals that if the value of $\nu$ increases, it means the corresponding structure spends much more power on redundant hardware. This is because the value of PE utilization never equals zero except the circuit is idle. Consequently, the smaller $\nu$ shows that the power consumption mostly results from the PE array. Likewise, higher PE utilization represents the well-allocated hardware, which indicates more PEs are conducting MAC operations. In other words, to reduce the power consumption of a CNN accelerator, it's necessary to reduce the scale of the PE array, which meets the target of this paper.\par

Moreover, function~(\ref{NU}) can also be computed under single convolution layers. This paper emphasizes the operation conditions of series and parallel CNN models, function~(\ref{NU}) also provides $\nu$ under different environments by adjusting the value of $N$ and $P_R$ in equation~(\ref{NU}).

\section{Experimental results}
The proposed SF-MMCN is implemented in Verilog HDL and logic synthesis with Design Compiler by Synopsys and implemented in TSMC 40nm technology under 400MHz clock frequency with Innovus by Candense. The bit-width of weight, input images data, and bias data are set to 16 bits fixed point. The proposed SF-MMCN can execute on VGG-16 and ResNet-18 in this paper. Figure~\ref{fig:SF-MMCN} is the implemented architecture of the proposed SF-MMCN. According to Figure~\ref{fig:SF-MMCN} there are 8 SF-MMCNs in implemented architecture.

\subsection{Selection of number of SF-MMCN}
\begin{figure}[H]
\centering
\includegraphics[width=0.4\textwidth]{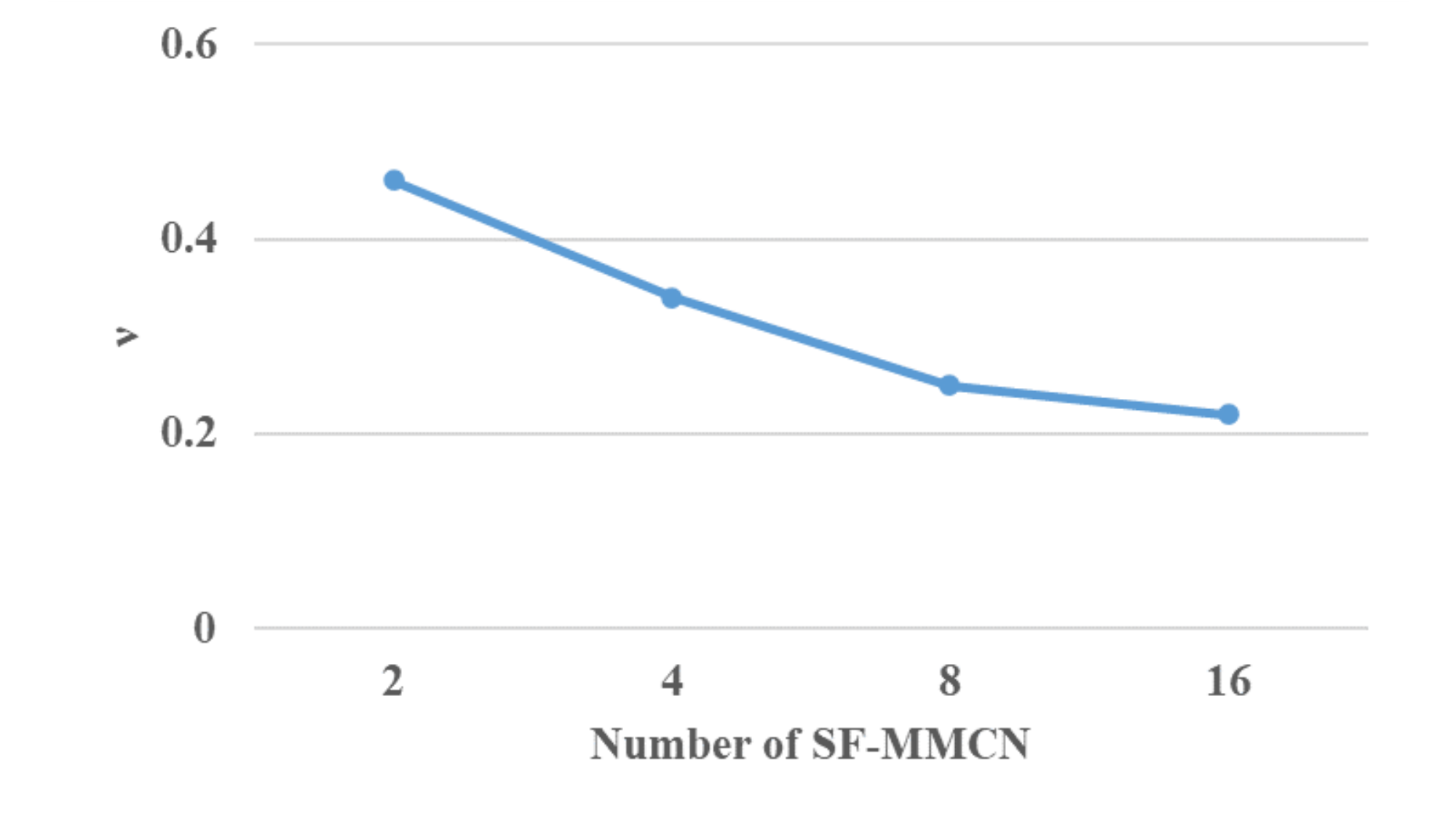}
\caption{The number of SF-MMCN and corresponding Efficiency Factor $\nu$.} 
\label{fig:HowSFMMCNNo.}
\end{figure}

As mentioned above, there are 8 proposed SF-MMCN in implemented structure. The reason why adopting 8 SF-MMCN is owing to the performance indicated in Figure~\ref{fig:HowSFMMCNNo.}. According to Figure~\ref{fig:HowSFMMCNNo.}, since the efficiency factor is the ratio between power and the actual executed PE in implementation, 8 SF-MMCNs perform great $\nu$ values. Although 16 SF-MMCNs provide the best $\nu$ value, the power consumption and the number of PEs become as large as the target of this paper. In other words, 16 SF-MMCNs shows the best result of $\nu$, but perform worse efficiency (GOPs/W). On the other hand, 2 and 4 SF-MMCNs provide the unwilling value of $\nu$ because a small MAC core unbalances the distribution of each hierarchy. 

\subsection{Results and Analysis}

\begin{figure}[H]
\centering
\subfloat[]{
\includegraphics[width=0.45\textwidth]{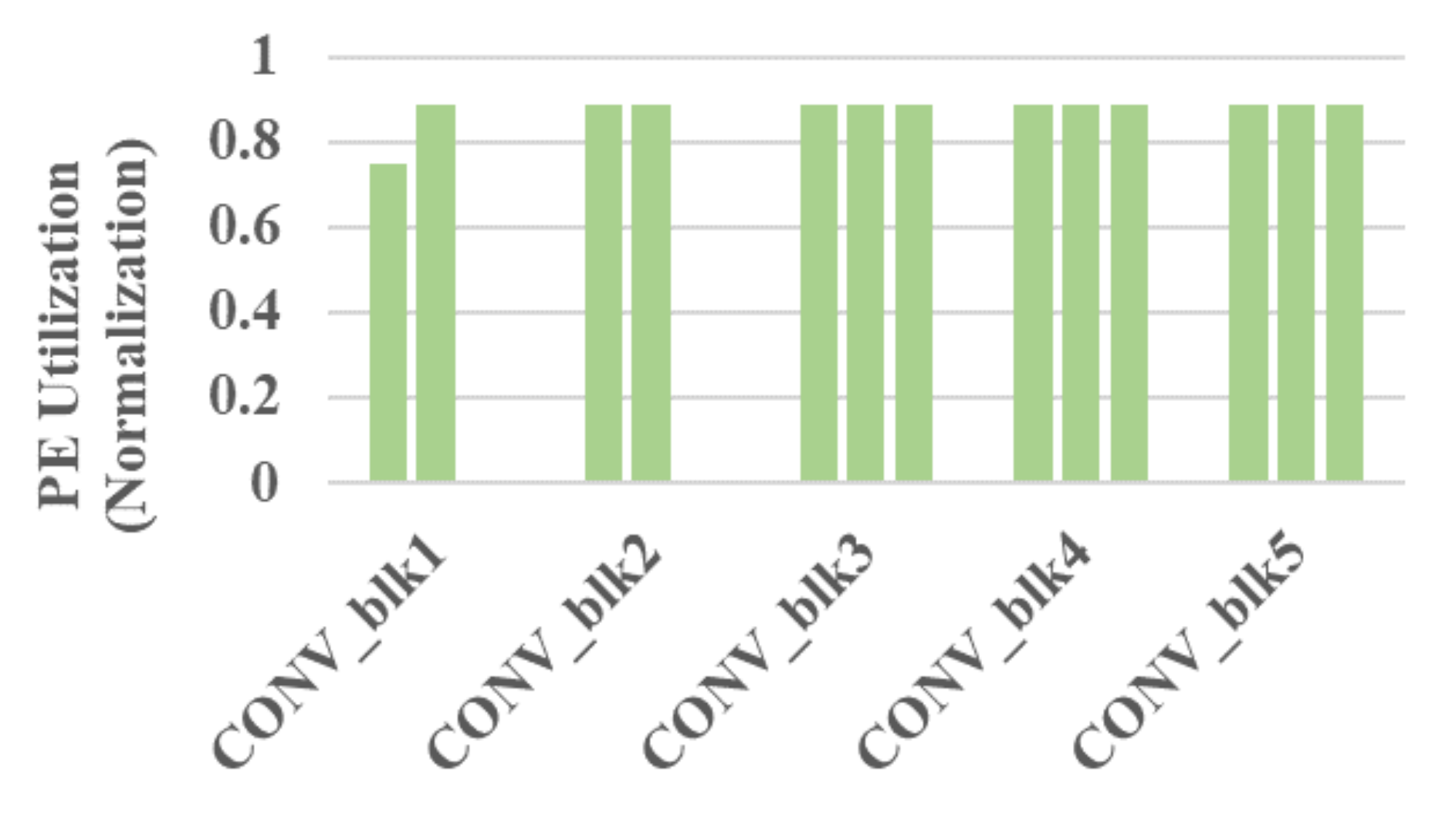}}
\\
\subfloat[]{
\includegraphics[width=0.45\textwidth]{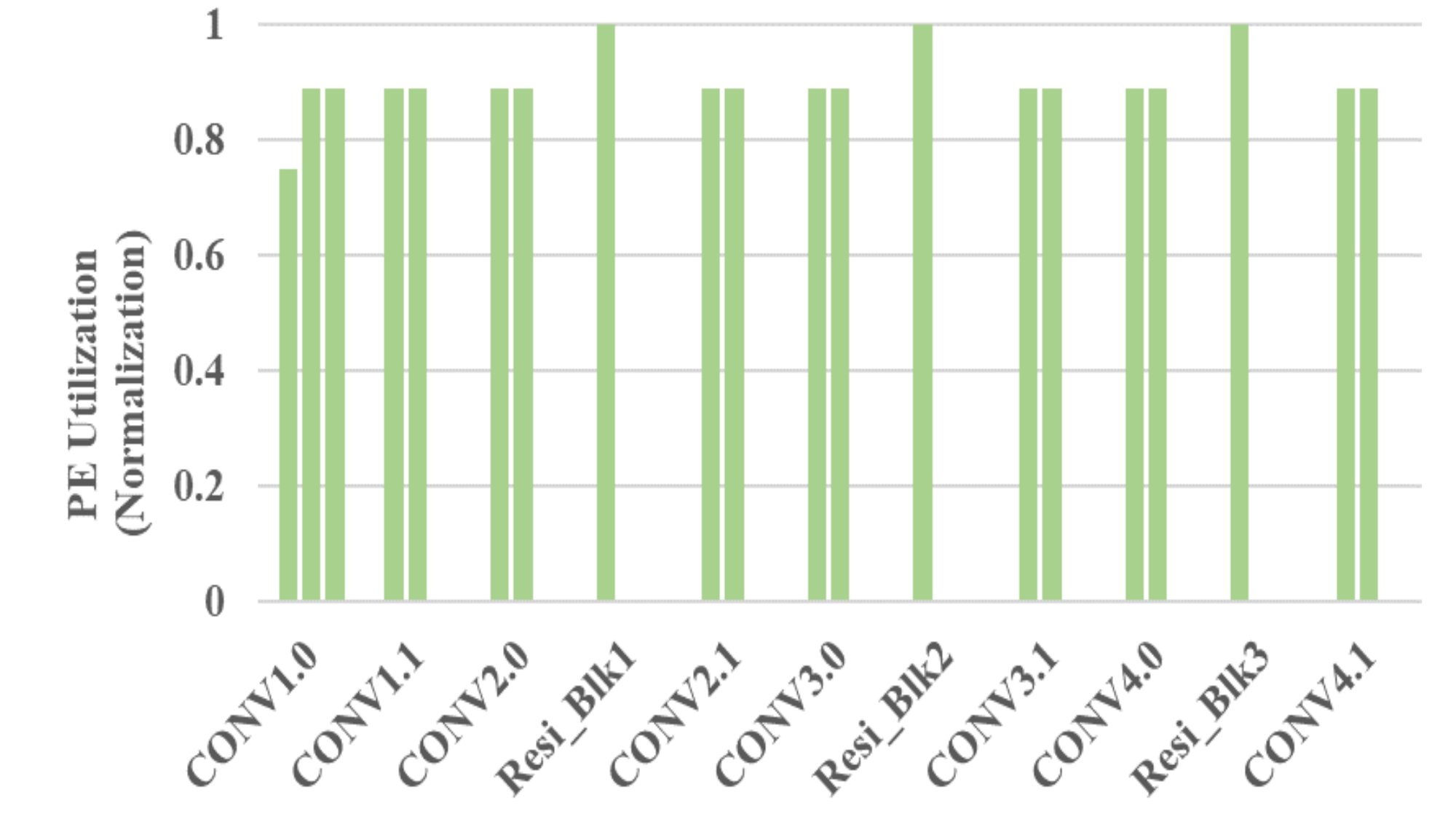}}
\caption{The PE utilization of the proposed SF-MMCN on (a) VGG-16. (b) ResNet-18.} 
\label{fig:PEU_vgg16_resnet18}
\end{figure}

\begin{table*}[htbp]
    \renewcommand\arraystretch{1.5}
    \caption{Comparison with Other Accelerators}
    \label{tab:final_comp}
    \centering
    {\begin{tabular}{lccccccc}
\hline \textbf{Performance}   &\textbf{TCASI'21\cite{CARLA}}  &\textbf{TCASI'21\cite{IECA}}   &\textbf{TCASI'22\cite{mem_effi}}   &\textbf{ISSCC'21\cite{ISSCC}}  &\textbf{ISSCC'23\cite{ISSCC23}}    &\textbf{MMCN\cite{my}}     &\textbf{This work}  \\ \hline    Frequency (MHz)             &200                    &250                            &700                                &100-470                        &20-400                             &200                        &400\\ 
\hline    Technology                  &65nm                   &55nm                           &28nm                               &28nm                           &28nm                               &90nm                       &40nm\\ 
\hline    Area ($mm^2$)               &6.2                    &2.75                           &NA                                 &1.9                            &7.29                               &0.36(core)                 &1.9\\ 
\hline    Gate count (NAND2)          &938k                   &NA                             &1.12M                              &NA                             &NA                                 &NA                         &211k\\ 
\hline    Precision (Bits)            &16                     &16                             &16                                 &8                              &1-8                                &16                         &16\\ 
\hline    Number of PEs               &196                    &168                            &288                                &144                            &8                                  &32                         &72\\ 
\hline    CNN models    &\makecell[c]{VGG-16\\ResNet-50} &\makecell[c]{VGG-16\\AlexNet} &VGG-16 &\makecell[c]{AlexNet/VGGNet\\GoogleNet/ResNet} &\makecell[c]{Eff.N-L0\\ViT-T/M.Mxr-B}                &VGG-16            &\makecell[c]{VGG-16\\ResNet-18}\\ 
\hline    Power (mW)                  &247                    &114.6                          &186.6                              &19.4 - 131.6                   &2.06-231.7                         &3.58(core)                 &18\\
\hline    Throughput (GOPs)           &77.4/75.4              &84.0                           &403                                &NA                             &1870-18900                         &2572.184                   &437.9\\ 
\hline    Energy Effi. (GOPs/W)       &0.31k/0.3k             &NA                             &2.1k                               &12.1k                          &907k-551k                          &718k                       &24.3k\\ 
\hline    Area Effi.  ($Gops/{mm^2}$) &12.48                   &30.55                          &NA                                 &745.1                          &720-2600                           &NA                        &230.47\\ 
\hline    Effi. actor ($\nu$)         &82.3                   &-                              &0.64                               &-                              &-                                  &0.11                       &0.02\\
\hline
    \end{tabular}}
\end{table*}

\subsubsection{PE utilization}
The target of this paper is to reduce the scale of the PE array and increase the utilization of PEs in hardware implementation. Figure~\ref{fig:PEU_vgg16_resnet18} (a) and Figure~\ref{fig:PEU_vgg16_resnet18} (b) are the performance of PE utilization on VGG-16 and ResNet-18 respectively. In the first layer of VGG-16, the PE utilization is lower than any other layer owing to the input image channel being 3. Therefore, only 6 of the proposed SF-MMCN are set to execute the convolution. In the rest layer of the VGG-16, the PE utilization is about 89\% because of the series structure in VGG-16. The PE\_9 in each of the proposed SF-MMCNs only processes the data reuse function instead of MAC operations.\par

On ResNet-18, Figure~\ref{fig:PEU_vgg16_resnet18} (b), the reason for PE utilization of the first layer is lower than the rest of the layers is the same as VGG-16. The series structure in ResNet-18 is also about 89\%. The highlight of ResNet-18 is residual structures. Due to the MAC operations and data transmission of the residual block, the PE utilization is up to 100\%. The definition of PE utilization in this paper is that all sub-circuits in PE are fully operated MAC computations in a single cycle of convolution. Therefore, although PE utilization in the first of VGG-16 and ResNet-18 are the lowest in whole CNN models, the density of MAC operation are much higher than CARLA~\cite{CARLA}, which only executes 3 PEs per cycle. 

\begin{table}[H]
    \renewcommand\arraystretch{1.25}
    \caption{Operation Efficiency Comparison}
    \label{Oper_effi}
    \centering
    \begin{tabular}{lccccc}
    \toprule
    \multirow{2}{*}{\textbf{Pixel}} &
        \multicolumn{2}{c}{\textbf{Cycles/CONV}} &
        \multicolumn{2}{c}{\textbf{No. of MAC}}  &
        \textbf{Speedup} \\
        & {\cite{CARLA}} & {SF-MMCN} & {\cite{CARLA}} & {SF-MMCN} & {(Normalization)}\\
        \midrule
        28 & 84  & 9 & 28  & 75  &$\times2.67$ \\
        32 & 96  & 9 & 32  & 85  &$\times2.67$ \\
       224 & 672 & 9 & 224 & 597 &$\times2.67$ \\
       \bottomrule
    \end{tabular}
\end{table}

\subsubsection{Operation Efficiency}

The PEs in every proposed SF-MMCN are executing MAC operations in parallel. Therefore, all PEs deliver convolution outputs at the same time after 9 cycles, which means an SF-MMCN can finish 8 convolution outputs in 9 cycles. Table~\ref{Oper_effi} is the operation efficiency comparison between CARLA~\cite{CARLA} and the proposed SF-MMCN. CARLA operates convolutions by each row of filters. Therefore, if the pixel of the input feature and the filter size are 28 and $3\times3$ respectively, CARLA has to spend around 3 times of pixel cycles to finish a convolution computation. Meanwhile, the MAC operations of the proposed SF-MMCN is about $\times2.67$ than CARLA due to the density of MAC operation in SF-MMCN is almost 100\% in single convolution computation.

\begin{figure}[H]
\centering
\includegraphics[width=0.45\textwidth]{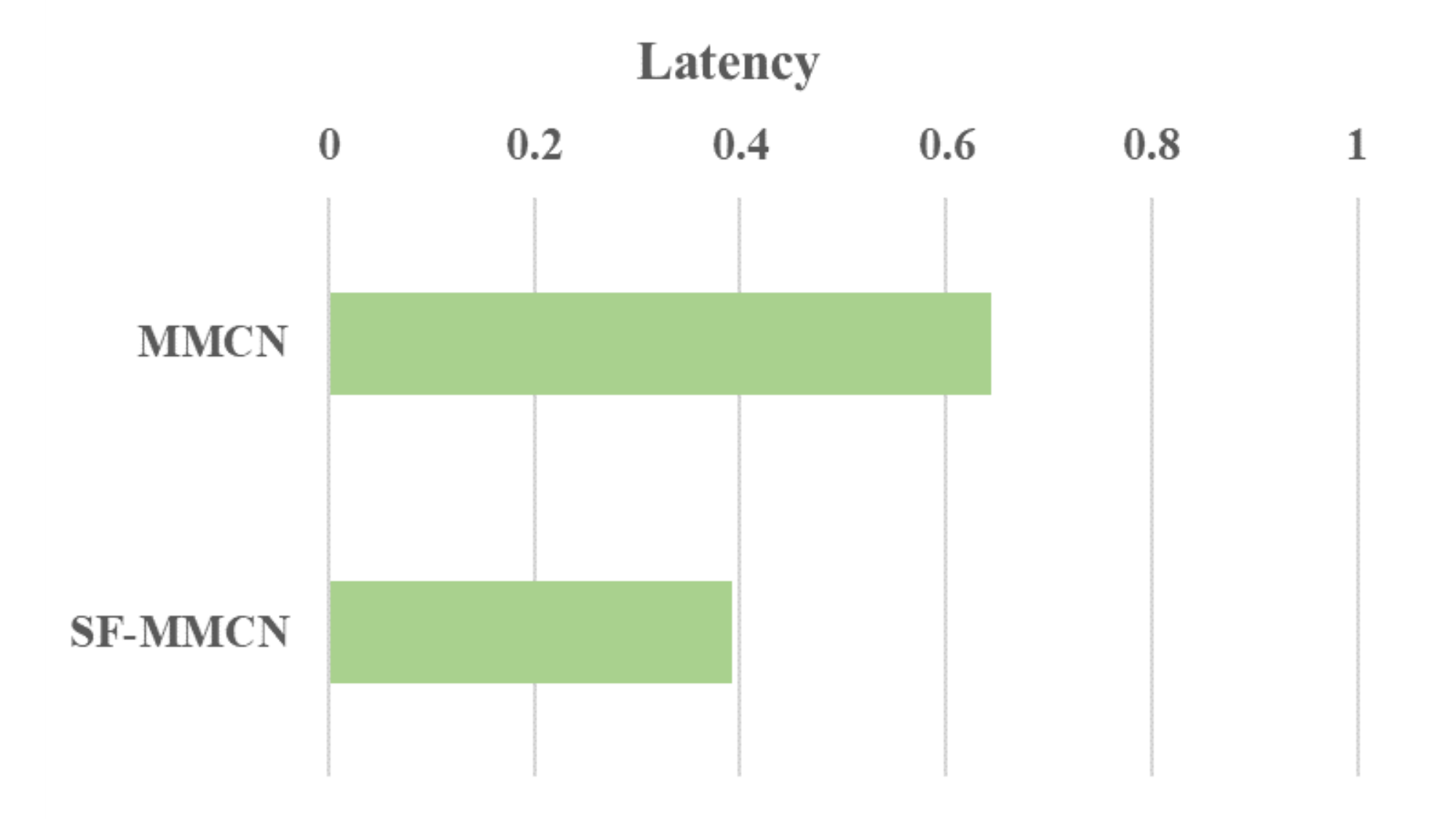}
\caption{The MAC operation  performance of the proposed SF-MMCN.} 
\label{fig:dataflow_sameW}
\end{figure}

\begin{figure}[H]
\centering
\includegraphics[width=0.45\textwidth]{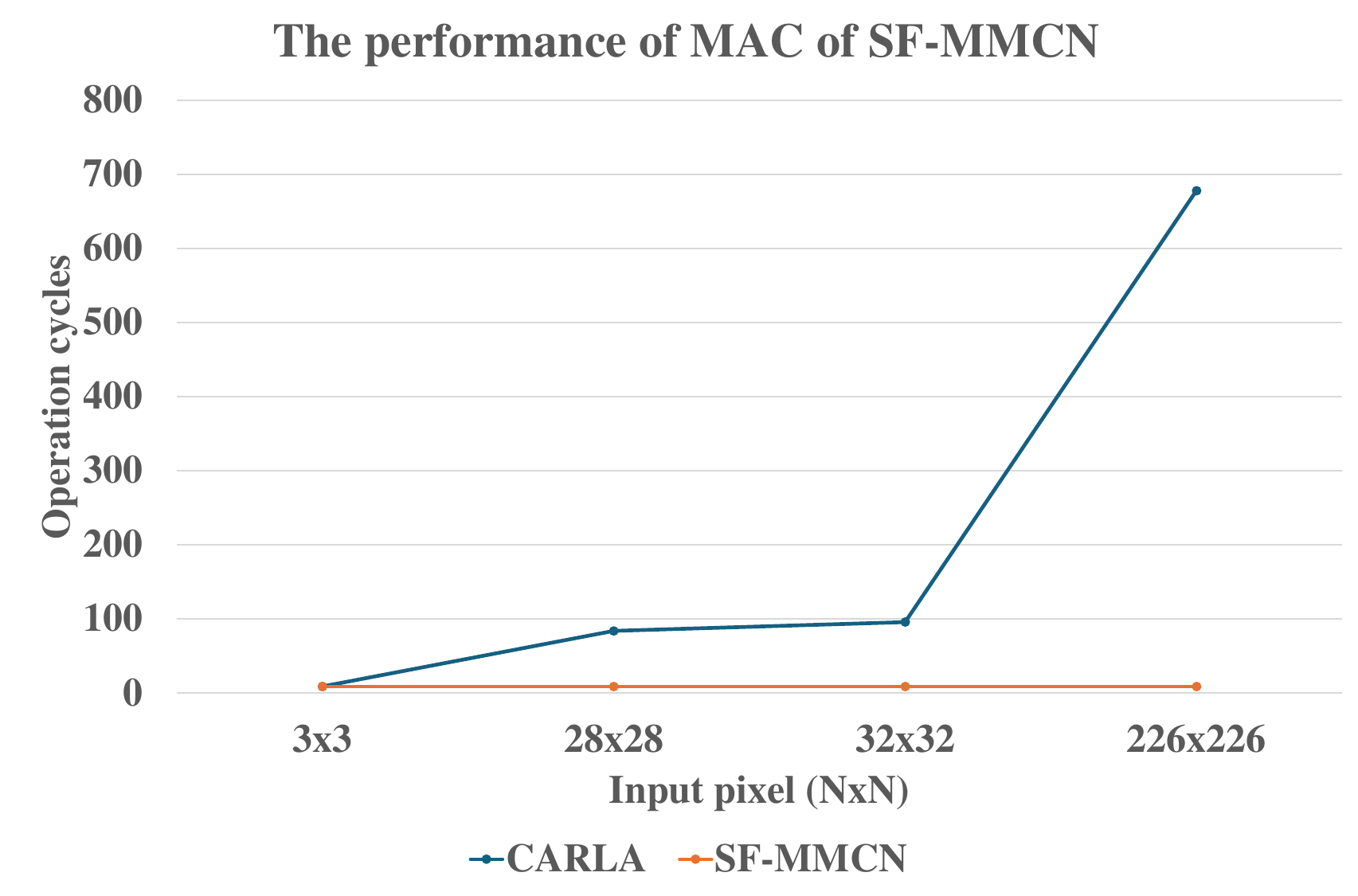}
\caption{The efficiency performance of the proposed SF-MMCN.} 
\label{fig:dataflow_diffW}
\end{figure}

To analyze the operation efficiency of the proposed SF-MMCN, Figure~\ref{fig:dataflow_sameW} and Figure~\ref{fig:dataflow_diffW} indicate the performance of the proposed structure under same and different pixels of input feature map and weight data. When computing under the same size of weight data, it's apparent that the number of cycles to deliver the first MAC output of the proposed SF-MMCN maintains only 9 despite the increasing of input data Figure~\ref{fig:dataflow_sameW}. On the other hand, CARLA~\cite{CARLA} requires additional cycles due to a modified low-power structure. The smallest cycle for CARLA has always been 3 times N according to Figure~\ref{fig:dataflow_sameW}. Once the size of input data exceeds 32, CARLA~\cite{CARLA} consumes many more cycles finishing convolution.\par

However, the size of weights is not always the same owing to the more complex CNN models coming out. Therefore, the analysis of the corresponding circumstance is revealed in Figure~\ref{fig:dataflow_diffW}. Where Wh and Ww represent the height and width of weight data respectively. The result shows that though the proposed structure requires to complete whole weight pixel in one convolution operation, the self-computing characteristic of single PE makes SF-MMCN deliver 9 convolution outputs in these cycles. However, CARLA~\cite{CARLA} only provides one convolution output in the same cycle on account of computing input image per row. Consequently, combined with Table~\ref{Oper_effi} and Figure~\ref{fig:dataflow_diffW}, the result shows that the proposed SF-MMCN can keep in high operation efficiency under different environments.

\begin{figure}[H]
\centering
\includegraphics[width=0.45\textwidth]{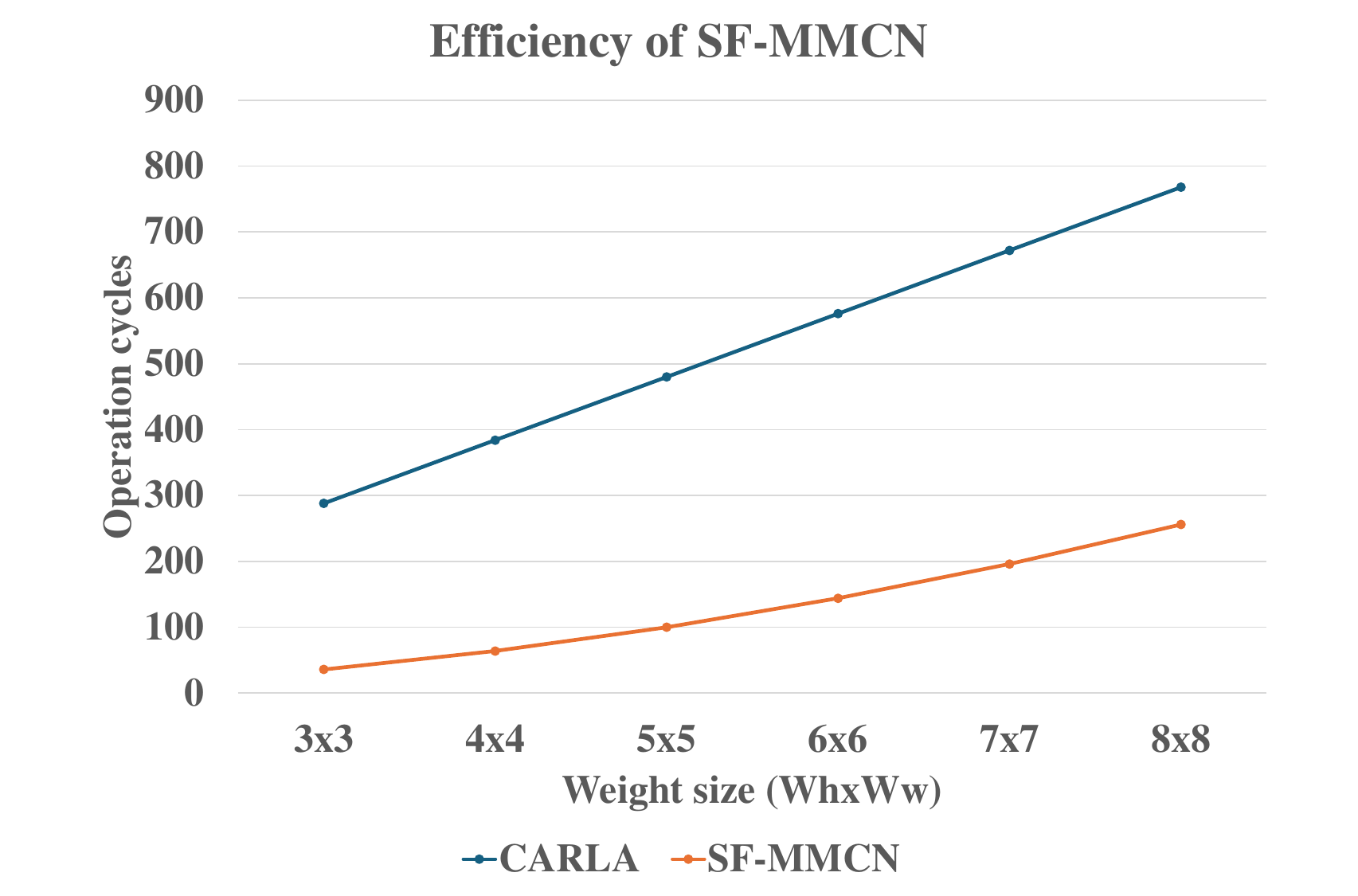}
\caption{Latency comparison between MMCN~\cite{my} and the proposed-SF-MMCN.} 
\label{fig:MMCNvsSFMMCN}
\end{figure}

\begin{figure}[H]
\centering
\includegraphics[width=0.5\textwidth]{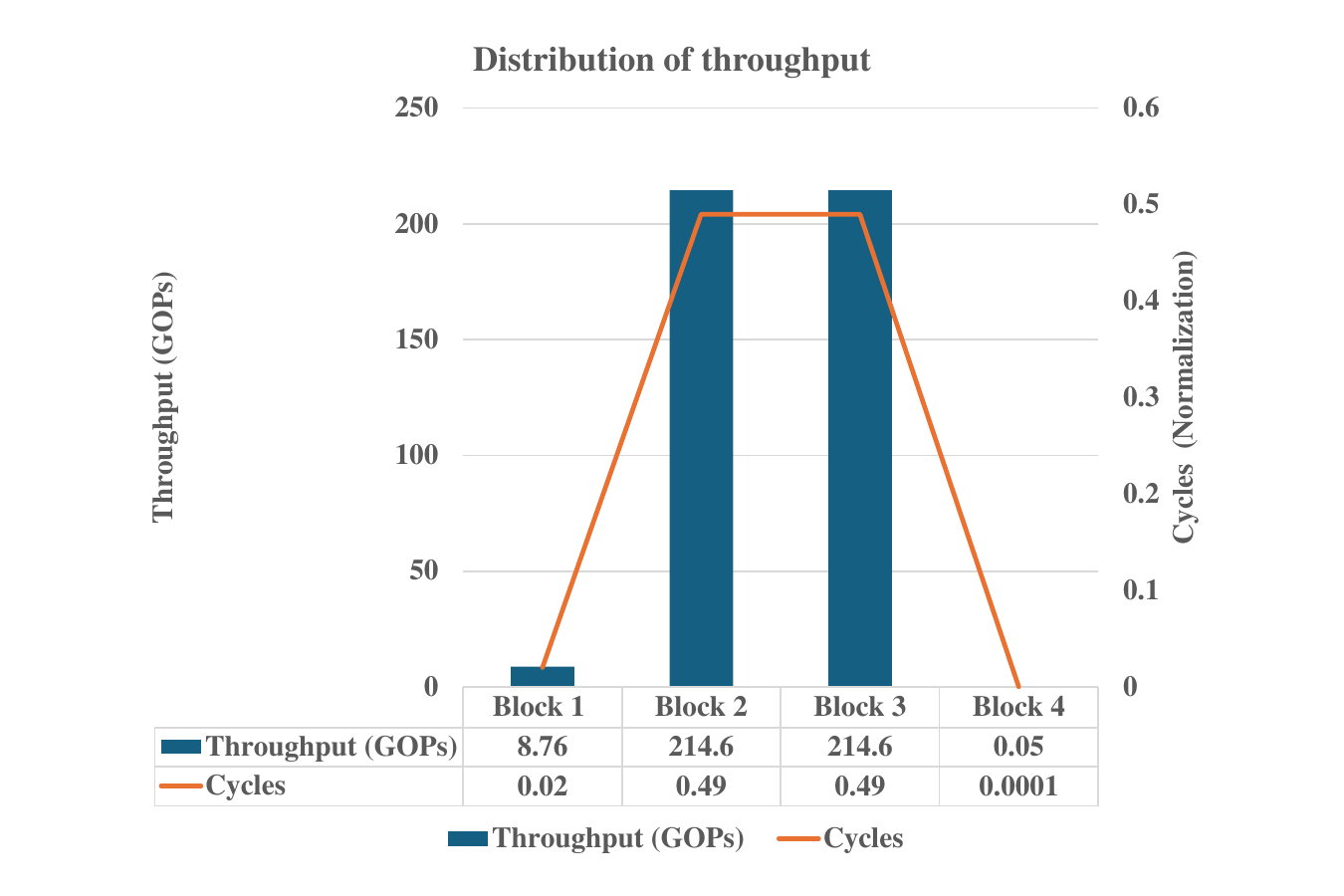}
\caption{The throughput performance of the proposed SF-MMCN on U-net.} 
\label{fig:through}
\end{figure}

The highlight of the implementation of the proposed SF-MMCN on diffusion model is the throughput performance. Since the complexity of a single block in U-net 
is much more than a single convolution layer in VGG or ResNet. Therefore, the best case of throughput is Block 2 and Block 3 according to Figure~\ref{fig:blk_structure} and Table~\ref{fig:through}. These are two of the most massive layers in a block in U-net. Since Block 1 and Block 4 are only for single channels or simple computation, which result in light percentages of operation cycles in one block in U-net. Therefore, if considering 4 blocks together to observe the throughput of a single block, it will achieve up to 437.976 GOPs. This result indicates that SF-MMCN successfully supports the diffusion model with high efficiency.

\subsection{Comparison with Other Accelerators}

The comparison of the proposed SF-MMCN and other remarkable accelerators is indicated in Table~\ref{tab:final_comp}. Although the proposed SF-MMCN only implemented VGG-16 and ResNet-18, it also supports other CNN models due to SF structure. This paper selects these models as examples of series and parallel models. The power and area performances are from synthesis results. First of all, as one of the targets of this study, the proposed SF-MMCN only required 72 PEs. The reason why \cite{ISSCC23} only adopted 8 PEs is that the specific structure of PEs in \cite{ISSCC23}. The PE in \cite{ISSCC23} not only consists of multipliers and accumulators but also data management circuits. What's more, though \cite{my} uses 32 PE in implementation, the latency of MMCN~\cite{my} on CNN models with parallel structure is not as well as the proposed SF-MMCN, Figure~\ref{fig:MMCNvsSFMMCN}. Due to the quantified MAC operations in the QNAP structure, the PE structure is different from traditional PE structures. Therefore, although the PE utilization in QNAP~\cite{ISSCC} is almost 100\%, the density of hardware executing reports lower hardware utilization. \par

The throughput of each accelerator in Table~\ref{tab:final_comp} is the peak throughput. Though MMCN~\cite{my} is the highest among all accelerators because of the different definitions of throughput. The throughput OPs in this paper are almost equal to FLOPs. Therefore, the proposed SF-MMCN performs a throughput of around 437.9 GOPs. Due to a large number of parameters and multiple layers in U-net, the proposed SF-MMCN achieves throughput incredibly compared to the old version MMCN~\cite{my}.

With high throughput performances of the proposed design, the other representative FoMs are operation efficiency $GOPs/W$ and area efficiency $GOPs/{mm^2}$. The former indicates the performance of the throughput of every power consumption unit of the accelerator, the latter is introduced in terms of the ratio of the throughput and area. According to Table~\ref{tab:final_comp}, all the information of the other state-of-the-art CNN accelerators is the result shown in corresponding references. Consequently, the operation efficiency of the proposed design is up to 81 times compared to CARLA~\cite{CARLA}. In area efficiency, the proposed structure advances nearly by 18 times (18.42).

The other factor of performance in this paper is $\nu$. As the index of operating density of hardware, the proposed SF-MMCN provides the smallest value of $\nu$ among other accelerators. Owing the specific PE structures in \cite{IECA}, \cite{ISSCC} and \cite{ISSCC23}, it's hard to receive the clear value of PE utilization ($U\_PE$ in equation~(\ref{UPE})). Therefore, the $\nu$ in Table~\ref{tab:final_comp} are "-". According to Table~\ref{tab:final_comp}, CARLA~\cite{CARLA} reports the highest $\nu$ value is because only 3 PEs operates in cycles. The $P_{act}$ only 3 when the size of the filter is $3\times3$. Consequently, the value of $\nu$ on the proposed SF-MMCN is the smallest representing that the proposed SF-MMCN can operate Neural Network computation without any redundant hardware, which reduces the power significantly.

\begin{figure}[H]
\centering
\includegraphics[width=0.45\textwidth]{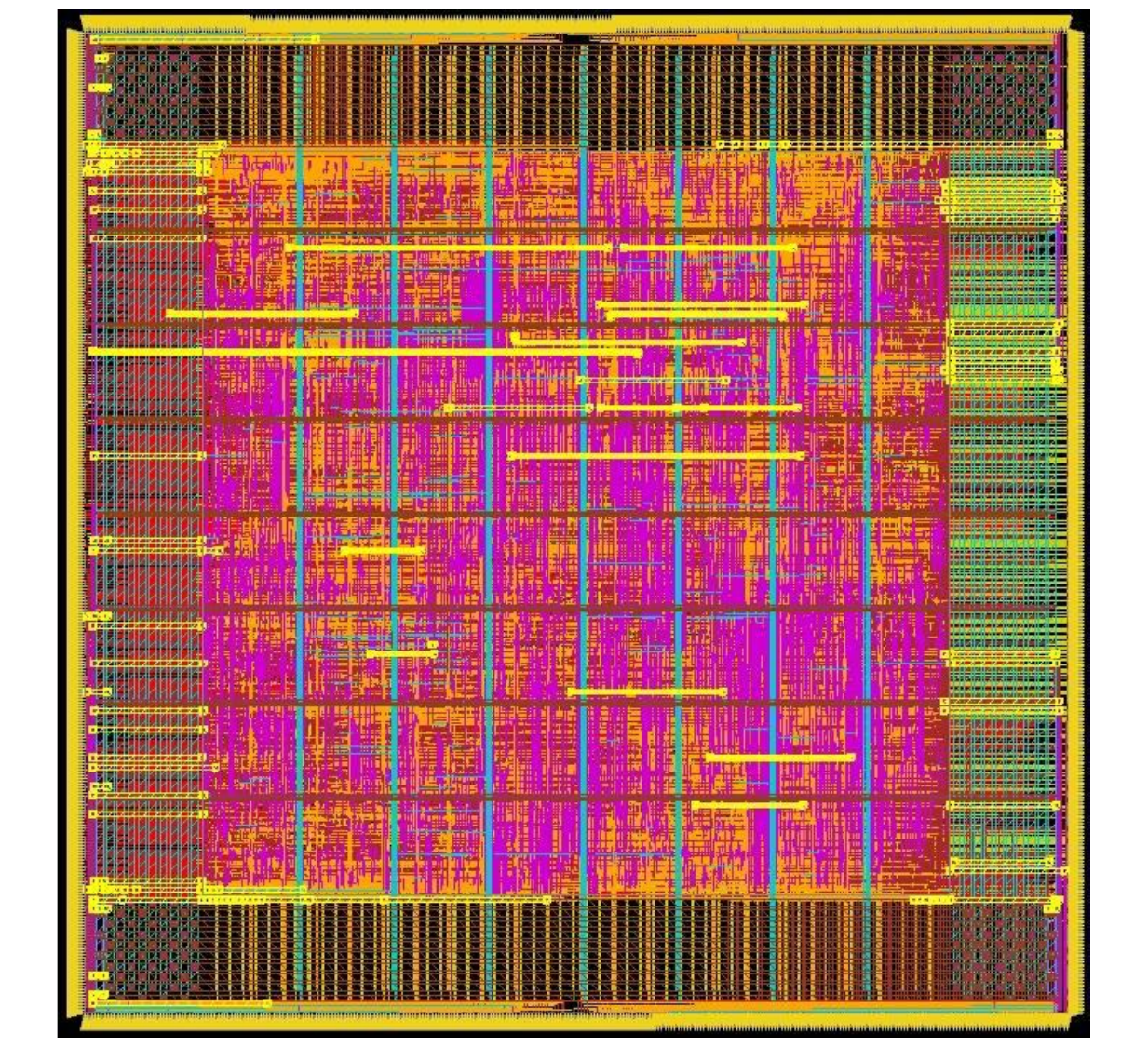}
\caption{The layout of the proposed SF-MMCN chip.} 
\label{fig:die}
\end{figure}

\subsection{Final Implementation}

\begin{table}[H]
    \renewcommand\arraystretch{1.5}
    \caption{Performance of the proposed SF-MMCN Chip}
    \label{tab:my_chip}
    \centering
    \begin{tabular}{lc}
    \toprule
         Performance &Specifications  \\
    \midrule
         Technology  &TSMC 40 nm     \\
\hline   Frequency   &200 MHz        \\
\hline   Voltage     &0.9 V     \\
\hline   Bit-width   &16 bits      \\
\hline   Chip size   &0.63mm $\times$ 0.63mm      \\
\hline   Chip area (Core)   &0.39 $mm^2$       \\
\hline   Total Power    &116.7 mW       \\
\hline   Efficiency     &3.75 $GOPs/mW$       \\
\hline   Area Effi.     &3752.36 $GOPs/{mm^2}$       \\
\bottomrule
    \end{tabular}
\end{table}

The layout depicted in Figure~\ref{fig:die} represents the proposed SF-MMCN chip design. Upon completing the placement and routing processes, the final performance metrics of the chip are summarized in Table~\ref{tab:my_chip}. Notably, the proposed SF-MMCN chip exhibits remarkable efficiency, reaching up to 3.75 ${GOPs}/{mW}$ while occupying a mere 0.17 ${mm^2}$ of hardware area. The throughput achieved is attributed to the operation of eight SF-MMCN cores. Scaling up the implementation by incorporating additional SF-MMCN cores promises a significant boost in throughput, facilitated by self-convolution by a processing element (PE). Furthermore, the compact hardware footprint suggests the potential for scalability without incurring substantial increases in hardware and power consumption. In conclusion, the proposed SF-MMCN chip successfully meets the objectives outlined in the thesis.

\section{Conclusion}

This paper provides the structure of SF-MMCN, a CNN accelerator with SF structure which lets SF-MMCN perform high throughput and acceleration of not only series but also parallel CNN structure. SF structure in the proposed SF-MMCN can operate convolution and conduct data reuse. SF-MMCN performs 18~mW of power consumption with only 1.9~$mm^2$ hardware area. The number of PE in SF-MMCN implementation is only 72, which is lower than other accelerators with traditional PE structures. The concept of efficient factor $\nu$ is also proposed in this paper, which represents the utilization density of PEs in an accelerator. Finally, SF-MMCN achieves energy efficiency 24.3~k($GOPs/W$) and 230.47($GOPS/mm^2$) of area efficiency.

\section*{Acknowledgments}
We would like to thank SUTD-ZJU IDEA Visiting Professor Grant (SUTD-ZJU (VP) 202103, and SUTD-ZJU Thematic Research Grant (SUTD-ZJU (TR) 202204), for supporting this work.


 
%

\bibliography{reference}

\begin{thebibliography}{10}
\providecommand{\url}[1]{#1}
\csname url@samestyle\endcsname
\providecommand{\newblock}{\relax}
\providecommand{\bibinfo}[2]{#2}
\providecommand{\BIBentrySTDinterwordspacing}{\spaceskip=0pt\relax}
\providecommand{\BIBentryALTinterwordstretchfactor}{4}
\providecommand{\BIBentryALTinterwordspacing}{\spaceskip=\fontdimen2\font plus
\BIBentryALTinterwordstretchfactor\fontdimen3\font minus \fontdimen4\font\relax}
\providecommand{\BIBforeignlanguage}[2]{{%
\expandafter\ifx\csname l@#1\endcsname\relax
\typeout{** WARNING: IEEEtran.bst: No hyphenation pattern has been}%
\typeout{** loaded for the language `#1'. Using the pattern for}%
\typeout{** the default language instead.}%
\else
\language=\csname l@#1\endcsname
\fi
#2}}
\providecommand{\BIBdecl}{\relax}
\BIBdecl

\bibitem{CarryapproFA}
M.~Ramasamy, G.~Narmadha, and S.~Deivasigamani, ``Carry based approximate full adder for low power approximate computing,'' in \emph{2019 7th International Conference on Smart Computing \& Communications (ICSCC)}.\hskip 1em plus 0.5em minus 0.4em\relax IEEE, 2019, pp. 1--4.

\bibitem{AppCons_LOA}
H.~R. Mahdiani, A.~Ahmadi, S.~M. Fakhraie, and C.~Lucas, ``Bio-inspired imprecise computational blocks for efficient vlsi implementation of soft-computing applications,'' \emph{IEEE Transactions on Circuits and Systems I: Regular Papers}, vol.~57, no.~4, pp. 850--862, 2009.

\bibitem{AppCons_LOCA}
A.~Dalloo, A.~Najafi, and A.~Garcia-Ortiz, ``Systematic design of an approximate adder: The optimized lower part constant-or adder,'' \emph{IEEE Transactions on Very Large Scale Integration (VLSI) Systems}, vol.~26, no.~8, pp. 1595--1599, 2018.

\bibitem{AppAddDNN}
S.~Raghuram and N.~Shashank, ``Approximate adders for deep neural network accelerators,'' in \emph{2022 35th International Conference on VLSI Design and 2022 21st International Conference on Embedded Systems (VLSID)}.\hskip 1em plus 0.5em minus 0.4em\relax IEEE, 2022, pp. 210--215.

\bibitem{PowerCarryPre}
P.~U. Joshi, D.~Khushlani, and R.~Khobragade, ``Power-area efficient computing technique for approximate multiplier with carry prediction,'' in \emph{2023 11th International Conference on Emerging Trends in Engineering \& Technology-Signal and Information Processing (ICETET-SIP)}.\hskip 1em plus 0.5em minus 0.4em\relax IEEE, 2023, pp. 1--4.

\bibitem{DRUM}
S.~Hashemi, R.~I. Bahar, and S.~Reda, ``Drum: A dynamic range unbiased multiplier for approximate applications,'' in \emph{2015 IEEE/ACM International Conference on Computer-Aided Design (ICCAD)}.\hskip 1em plus 0.5em minus 0.4em\relax IEEE, 2015, pp. 418--425.

\bibitem{SSM_DSM}
I.~Hammad, L.~Li, K.~El-Sankary, and W.~M. Snelgrove, ``Cnn inference using a preprocessing precision controller and approximate multipliers with various precisions,'' \emph{IEEE Access}, vol.~9, pp. 7220--7232, 2021.

\bibitem{BPE_CPE}
W.~Liu, J.~Lin, and Z.~Wang, ``A precision-scalable energy-efficient convolutional neural network accelerator,'' \emph{IEEE Transactions on Circuits and Systems I: Regular Papers}, vol.~67, no.~10, pp. 3484--3497, 2020.

\bibitem{Comp_basic}
J.~Tonfat and R.~Reis, ``Low power 3--2 and 4--2 adder compressors implemented using astran,'' in \emph{2012 IEEE 3rd Latin American Symposium on Circuits and Systems (LASCAS)}.\hskip 1em plus 0.5em minus 0.4em\relax IEEE, 2012, pp. 1--4.

\bibitem{TwoComp4_2}
L.~Sayadi, S.~Timarchi, and A.~Sheikh-Akbari, ``Two efficient approximate unsigned multipliers by developing new configuration for approximate 4: 2 compressors,'' \emph{IEEE Transactions on Circuits and Systems I: Regular Papers}, vol.~70, no.~4, pp. 1649--1659, 2023.

\bibitem{LowPowerComp4_2}
A.~G. Strollo, D.~De~Caro, E.~Napoli, N.~Petra, and G.~Di~Meo, ``Low-power approximate multiplier with error recovery using a new approximate 4-2 compressor,'' in \emph{2020 IEEE International Symposium on Circuits and Systems (ISCAS)}.\hskip 1em plus 0.5em minus 0.4em\relax IEEE, 2020, pp. 1--4.

\bibitem{Eyeriss}
Y.-H. Chen, T.~Krishna, J.~S. Emer, and V.~Sze, ``Eyeriss: An energy-efficient reconfigurable accelerator for deep convolutional neural networks,'' \emph{IEEE journal of solid-state circuits}, vol.~52, no.~1, pp. 127--138, 2016.

\bibitem{ReconvHA}
A.~Ansari, K.~Gunnam, and T.~Ogunfunmi, ``An efficient reconfigurable hardware accelerator for convolutional neural networks,'' in \emph{2017 51st Asilomar Conference on Signals, Systems, and Computers}.\hskip 1em plus 0.5em minus 0.4em\relax IEEE, 2017, pp. 1337--1341.

\bibitem{Flexible}
C.~Yang, H.~Zhang, X.~Wang, and L.~Geng, ``An energy-efficient and flexible accelerator based on reconfigurable computing for multiple deep convolutional neural networks,'' in \emph{2018 14th IEEE International Conference on Solid-State and Integrated Circuit Technology (ICSICT)}.\hskip 1em plus 0.5em minus 0.4em\relax IEEE, 2018, pp. 1--3.

\bibitem{CARLA}
M.~Ahmadi, S.~Vakili, and J.~P. Langlois, ``Carla: A convolution accelerator with a reconfigurable and low-energy architecture,'' \emph{IEEE Transactions on Circuits and Systems I: Regular Papers}, vol.~68, no.~8, pp. 3184--3196, 2021.

\bibitem{OverFF}
W.~Xu, Z.~Zhang, X.~You, and C.~Zhang, ``Reconfigurable and low-complexity accelerator for convolutional and generative networks over finite fields,'' \emph{IEEE Transactions on Computer-Aided Design of Integrated Circuits and Systems}, vol.~39, no.~12, pp. 4894--4907, 2020.

\bibitem{Pre-scalable}
W.~Liu, J.~Lin, and Z.~Wang, ``A precision-scalable energy-efficient convolutional neural network accelerator,'' \emph{IEEE Transactions on Circuits and Systems I: Regular Papers}, vol.~67, no.~10, pp. 3484--3497, 2020.

\bibitem{ECG}
J.~Lu, D.~Liu, A.~Hu, C.~Zhang, C.~Mo, R.~Guo, and H.~Li, ``A low-cost and configurable hardware architecture of sparse 1-d cnn for ecg classification,'' in \emph{2022 IEEE 16th International Conference on Solid-State \& Integrated Circuit Technology (ICSICT)}.\hskip 1em plus 0.5em minus 0.4em\relax IEEE, 2022, pp. 1--3.

\bibitem{ISSCC}
H.~Mo, W.~Zhu, W.~Hu, G.~Wang, Q.~Li, A.~Li, S.~Yin, S.~Wei, and L.~Liu, ``9.2 a 28nm 12.1 tops/w dual-mode cnn processor using effective-weight-based convolution and error-compensation-based prediction,'' in \emph{2021 IEEE International Solid-State Circuits Conference (ISSCC)}, vol.~64.\hskip 1em plus 0.5em minus 0.4em\relax IEEE, 2021, pp. 146--148.

\bibitem{ISCAS}
Y.~Zeng, H.~Sun, J.~Katto, and Y.~Fan, ``Accelerating convolutional neural network inference based on a reconfigurable sliced systolic array,'' in \emph{2021 IEEE International Symposium on Circuits and Systems (ISCAS)}.\hskip 1em plus 0.5em minus 0.4em\relax IEEE, 2021, pp. 1--5.

\bibitem{TCAD}
W.~Xu, Z.~Zhang, X.~You, and C.~Zhang, ``Reconfigurable and low-complexity accelerator for convolutional and generative networks over finite fields,'' \emph{IEEE Transactions on Computer-Aided Design of Integrated Circuits and Systems}, vol.~39, no.~12, pp. 4894--4907, 2020.

\bibitem{diffusion}
J.~Ho, A.~Jain, and P.~Abbeel, ``Denoising diffusion probabilistic models,'' \emph{Advances in neural information processing systems}, vol.~33, pp. 6840--6851, 2020.

\bibitem{unet}
O.~Ronneberger, P.~Fischer, and T.~Brox, ``U-net: Convolutional networks for biomedical image segmentation,'' in \emph{Medical image computing and computer-assisted intervention--MICCAI 2015: 18th international conference, Munich, Germany, October 5-9, 2015, proceedings, part III 18}.\hskip 1em plus 0.5em minus 0.4em\relax Springer, 2015, pp. 234--241.

\bibitem{my}
H.-K. Hsu, I.-C. Wey, and T.~H. Teo, ``A energy-efficient re-configurable multi-mode convolution neuron network accelerator,'' in \emph{2023 IEEE 16th International Symposium on Embedded Multicore/Many-core Systems-on-Chip (MCSoC)}.\hskip 1em plus 0.5em minus 0.4em\relax IEEE, 2023, pp. 45--50.

\bibitem{hsu2024sfmmcn}
\BIBentryALTinterwordspacing
------, ``Sf-mmcn: A low power re-configurable server flow convolution neural network accelerator,'' 2024. [Online]. Available: \url{https://arxiv.org/abs/2403.10542}
\BIBentrySTDinterwordspacing

\bibitem{pipe_resnet_1}
M.~Alwani, H.~Chen, M.~Ferdman, and P.~Milder, ``Fused-layer cnn accelerators,'' in \emph{2016 49th Annual IEEE/ACM International Symposium on Microarchitecture (MICRO)}.\hskip 1em plus 0.5em minus 0.4em\relax IEEE, 2016, pp. 1--12.

\bibitem{pipe_resnet_2}
H.~Mo, L.~Liu, W.~Zhu, Q.~Li, H.~Liu, S.~Yin, and S.~Wei, ``A multi-task hardwired accelerator for face detection and alignment,'' \emph{IEEE Transactions on Circuits and Systems for Video Technology}, vol.~30, no.~11, pp. 4284--4298, 2019.

\bibitem{IECA}
B.~Huang, Y.~Huan, H.~Chu, J.~Xu, L.~Liu, L.~Zheng, and Z.~Zou, ``Ieca: An in-execution configuration cnn accelerator with 30.55 gops/mm$^2$ area efficiency,'' \emph{IEEE Transactions on Circuits and Systems I: Regular Papers}, vol.~68, no.~11, pp. 4672--4685, 2021.

\bibitem{mem_effi}
Z.~Shao, X.~Chen, L.~Du, L.~Chen, Y.~Du, W.~Zhuang, H.~Wei, C.~Xie, and Z.~Wang, ``Memory-efficient cnn accelerator based on interlayer feature map compression,'' \emph{IEEE Transactions on Circuits and Systems I: Regular Papers}, vol.~69, no.~2, pp. 668--681, 2021.

\bibitem{ISSCC23}
S.~Moon, H.-G. Mun, H.~Son, and J.-Y. Sim, ``A 127.8 tops/w arbitrarily quantized 1-to-8b scalable-precision accelerator for general-purpose deep learning with reduction of storage, logic and latency waste,'' in \emph{2023 IEEE International Solid-State Circuits Conference (ISSCC)}.\hskip 1em plus 0.5em minus 0.4em\relax IEEE, 2023, pp. 21--23.

\end{thebibliography}
\bibliographystyle{IEEEtran}
\end{document}